# Self-Supervised Learning of Physics-Guided Reconstruction Neural Networks without Fully-Sampled Reference Data

Burhaneddin Yaman[1,2], Seyed Amir Hossein Hosseini[1,2], Steen Moeller[2], Jutta Ellermann[2], Kâmil Uğurbil[2], Mehmet Akçakaya[1,2],

[1]Department of Electrical and Computer Engineering, and [2]Center for Magnetic Resonance Research, University of Minnesota, Minneapolis, MN

**Correspondence to:**

Mehmet Akçakaya, Ph. D.

University of Minnesota, 200 Union Street S.E., Minneapolis, MN, 55455

Phone: 612-625-1343; Fax: 612-625-4583

E-mail: akcakaya@umn.edu

**Funding:**

NIH, Grant numbers: U01EB025144, P41EB027061; NSF, Grant number: CAREER CCF-1651825

**Running Head:** Self-supervised Learning of Physics-Guided Neural Network

# Abstract

**Purpose:** To develop a strategy for training a physics-guided MRI reconstruction neural network without a database of fully-sampled datasets.

**Theory and Methods:** Self-supervised learning via data under-sampling (SSDU) for physics-guided deep learning (DL) reconstruction partitions available measurements into two disjoint sets, one of which is used in the data consistency units in the unrolled network and the other is used to define the loss for training. The proposed training without fully-sampled data is compared to fully-supervised training with ground-truth data, as well as conventional compressed sensing and parallel imaging methods using the publicly available fastMRI knee database. The same physics-guided neural network is used for both proposed SSDU and supervised training. The SSDU training is also applied to prospectively 2-fold accelerated high-resolution brain datasets at different acceleration rates, and compared to parallel imaging.

**Results:** Results on five different knee sequences at acceleration rate of 4 shows that proposed self-supervised approach performs closely with supervised learning, while significantly outperforming conventional compressed sensing and parallel imaging, as characterized by quantitative metrics and a clinical reader study. The results on prospectively sub-sampled brain datasets, where supervised learning cannot be employed due to lack of ground-truth reference, show that the proposed self-supervised approach successfully perform reconstruction at high acceleration rates (4, 6 and 8). Image readings indicate improved visual reconstruction quality with the proposed approach compared to parallel imaging at acquisition acceleration.

**Conclusion:** The proposed SSDU approach allows training of physics-guided DL-MRI reconstruction without fully-sampled data, while achieving comparable results with supervised DL-MRI trained on fully-sampled data.

# Introduction

Data acquisition in MRI is inherently slow, necessitating the use of accelerated imaging techniques. In these approaches, data is acquired at sub-Nyquist rates, and reconstructed using additional information. Parallel imaging exploits the redundancies between receiver coils and is the most clinically used approach (1-3). Compressed sensing is another method that utilizes the compressibility of images based on linear sparsifying transforms for a regularized reconstruction (4-9), which can also be synergistically combined with multi-coil acquisitions (10-12). At high acceleration rates, parallel imaging suffers from noise amplification (13-15), while compressed sensing may lead to residual artifacts (16,17). Furthermore, compressed sensing reconstruction is computationally lengthy in nature and typically requires empirical fine-tuning of regularization parameters, although recent approaches using rapid self-tuning show promise for principled parameter selection (18,19).

Recently, deep learning (DL) has gained interest for high-quality accelerated MRI. DL based MRI reconstruction algorithms can be roughly divided into two categories, purely data-driven and physics-guided (20). In purely data-driven approaches, a mapping between the undersampled k-space/aliased image to full k-space/artifact-free image is learned (21-26). In the so-called physics-guided methods, the knowledge of the forward encoding operator, which contains the undersampling pattern and typically the coil sensitivities, is taken into account to solve an inverse problem based on a regularized least squares objective function (27-35). Some other works have directly worked with multi-coil data without explicitly including the coil sensitivities (36,37). These techniques unroll an iterative reconstruction algorithm for solving this objective method for a fixed number of iterations. The unrolled network alternates between data consistency and regularization, where the regularization is implemented implicitly using a neural network. Subsequently, these unrolled networks are trained end-to-end with a loss function that characterizes similarity with a reference image obtained from fully-sampled data (20). The parameters of the network can be different across the unrolled iterations (27,31) or shared across them (28,33).

The aforementioned physics-guided methods have been trained in a supervised manner, where fully-sampled data is used as a reference during the training. However, in many practical imaging scenarios, it is infeasible to acquire fully-sampled datasets. For instance, when imaging moving organs, such as the heart, there is often a short period of time during which the data needs to be acquired. Example acquisitions include real-time imaging, myocardial perfusion, and numerous contrast-enhanced scans (38-40). Another hindrance for fully-sampled acquisitions in some applications include the signal decay. This is pronounced in acquisitions, such as diffusion MRI with echo-planar imaging, where the signal decays quickly with $T_2^*$, thus prohibiting use of fully-sampled acquisitions especially at high resolutions (41,42). In several other scenarios such as whole-heart coronary MRI or high-resolution anatomical brain imaging, it is impractical to acquire fully-sampled datasets as the scan time becomes extremely lengthy.

Furthermore, accelerated imaging methods are often used to improve acquisition resolution. When higher acceleration rates are achievable, these are not solely used for image time reduction, but rather a trade-off is made with improved resolution (12,43,44). However, this newer resolution may necessitate re-training of the DL reconstruction, since neural networks do not necessarily generalize across different resolutions, as depicted in **Supporting Information Figure S1**. Thus, if fully-sampled data is required for training at higher resolutions, this may lead to excessive scan times, even for anatomical imaging protocols, making it difficult to make protocol changes to fully utilize the benefits of accelerated imaging.

In this study, we sought to develop a new self-supervised learning approach to train physics-guided DL-MRI reconstruction without fully-sampled reference data. The proposed self-supervised approach which we term as Self-Supervision via Data Undersampling (SSDU) splits the acquired k-space indices into two disjoint sets. One of these is used in the data consistency unit for the network, while the other set is used to define the loss function in k-space. Hence, end-to-end training and evaluation of the network is done through only the acquired measurements without making any other assumptions about image output or characteristics. We apply the proposed self-supervised training without fully-sampled data, on the fastMRI knee datasets and prospectively undersampled high-resolution brain MRI datasets. These are compared to parallel imaging, compressed sensing and a supervised training of a DL-MRI network when fully-sampled reference data is available. Our results indicate that the proposed self-supervised method performs similarly to the supervised approach trained on fully-sampled data, although it is trained only on undersampled data.

## Theory

### Physics-Guided Neural Networks for MRI Reconstruction

Let **x** denote the image to be recovered and $\mathbf{y}_\Omega$ represent acquired k-space measurements with undersampling pattern $\Omega$. The forward model for the acquisition is given as

$$\mathbf{y}_\Omega = \mathbf{E}_\Omega \mathbf{x} + \mathbf{n}, \qquad (1)$$

where $\mathbf{E}_\Omega: \mathbb{C}^{M_1 \times M_2} \to \mathbb{C}^P$ is the encoding operator including a partial Fourier matrix sampling the locations specified by $\Omega$ and the coil sensitivities, and $\mathbf{n} \in \mathbb{C}^P$ is measurement noise. The forward model presented in Equation [1] is usually ill-conditioned due to sub-Nyquist sampling and hence regularizers that induce prior information is incorporated into the objective function for the reconstruction. Possible choices for the regularizer include total variation (10,45,46), $\ell_1$-norm of wavelet coefficients (4,8,47), sparsity in adaptive transform domains (9,48), and more recently neural networks (27,28,33). The image recovery is then formulated as an optimization problem

$$\arg\min_{x} \|\mathbf{y}_\Omega - \mathbf{E}_\Omega \mathbf{x}\|_2^2 + \mathcal{R}(x), \tag{2}$$

where the first term represents data consistency with acquired measurements, while $\mathcal{R}(\cdot)$ is a regularization term. The optimization problem in Equation [2] can be solved in numerous ways, including proximal gradient descent, variable splitting with quadratic penalty, alternating direction method of multipliers among others (27,30,32,49). In this study, we will consider the variable splitting with quadratic penalty approach (50) for implementation, which has also been used in previous physics-guided DL-MRI approaches (28,32). In this method, data consistency and regularization are decoupled as

$$\arg\min_{x,z} \|\mathbf{y}_\Omega - \mathbf{E}_\Omega \mathbf{x}\|_2^2 + \mu \|\mathbf{x} - \mathbf{z}\|_2^2 + \mathcal{R}(\mathbf{z}), \tag{3}$$

where $\mathbf{z}$ is the auxiliary variable that is initially constrained to be equal to $\mathbf{x}$, and $\mu$ is the parameter for the quadratic penalty for relaxing this intermediate constrained problem to an unconstrained one. The optimization problem in Equation [3] is then solved iteratively by alternating the minimization over the variables $\mathbf{x}$ and $\mathbf{z}$ as follows

$$\mathbf{z}^{(i-1)} = \arg\min_{\mathbf{z}} \mu \|\mathbf{x}^{(i-1)} - \mathbf{z}\|_2^2 + \mathcal{R}(\mathbf{z}), \tag{4}$$

$$\mathbf{x}^{(i)} = \arg\min_{\mathbf{x}} \|\mathbf{y}_\Omega - \mathbf{E}_\Omega \mathbf{x}\|_2^2 + \mu \|\mathbf{x} - \mathbf{z}^{(i-1)}\|_2^2, \tag{5}$$

where $\mathbf{x}^{(0)}$ is the initial image obtained from zero-filled under-sampled k-space data, $\mathbf{x}^{(i)}$ is the network output at iteration $i$ and $\mathbf{z}^{(i)}$ is an intermediate variable. In compressed sensing methods, these problems are solved in an iterative manner by alternating between the regularizer and data consistency units until a stopping criterion met as shown in **Figure 1a**.

In physics-guided DL-MRI approaches, this iterative algorithm is unrolled for a fixed number of iterations, as depicted in **Figure 1b**. The regularization sub-problem in Equation [4] is implicitly solved using a neural network. The data consistency sub-problem in Equation [5] has a closed form solution

$$\mathbf{x}^{(i)} = (\mathbf{E}_\Omega^H \mathbf{E}_\Omega + \mu \mathbf{I})^{-1}(\mathbf{E}_\Omega^H \mathbf{y} + \mu \mathbf{z}^{(i-1)}), \tag{6}$$

where $\mathbf{I}$ is the identity operator and $(\cdot)^H$ is the conjugate transpose operator. Equation [6] can be solved using gradient descent or conjugate gradient, which itself is unrolled for a number of iterations (28).

## Supervised Training with Fully-Sampled Reference Datasets

Supervised learning performs end-to-end training using ground truth images as the reference labels for the training loss function (21,27). Ground truth images are obtained through SENSE-1 coil combination (2), which is the sum across the coil dimension of the product of the conjugate of the coil sensitivity maps with the corresponding coil images (31,32). Suppose that $\mathbf{x}_{ref}^i$ is the ground truth image for subject $i$, and $f(\mathbf{y}_\Omega^i, \mathbf{E}_\Omega^i; \boldsymbol{\theta})$ denotes the output of the unrolled network that is parametrized by $\boldsymbol{\theta}$ for subsampled k-space data $\mathbf{y}_\Omega^i$ and corresponding encoding matrix $\mathbf{E}_\Omega^i$ of the same subject $i$. The supervised training of a physics-guided DL-MRI method can be performed by minimizing the image domain loss

$$\min_{\boldsymbol{\theta}} \frac{1}{N} \sum_{i=1}^{N} \mathcal{L}\left(\mathbf{x}_{ref}^i, f(\mathbf{y}_\Omega^i, \mathbf{E}_\Omega^i; \boldsymbol{\theta})\right), \qquad (7)$$

where $N$ is the number of fully-sampled training data in the database, $\mathcal{L}(.,.)$ denotes the loss between the ground truth and network output image(27,28,31). Alternatively, supervised training may be evaluated in k-space as

$$\min_{\boldsymbol{\theta}} \frac{1}{N} \sum_{i=1}^{N} \mathcal{L}\left(\mathbf{y}_{ref}^i, \mathbf{E}_{full}^i \left(f(\mathbf{y}_\Omega^i, \mathbf{E}_\Omega^i; \boldsymbol{\theta})\right)\right), \qquad (8)$$

where $\mathbf{y}_{ref}^i$ is the fully-sampled reference k-space and $\mathbf{E}_{full}^i$ is the fully-sampled encoding operator that transforms network output to k-space across coils. Example loss functions include $\ell_1$ norm, $\ell_2$ norm, mixed norm and perception based loss(25,32,51-53). We note that the subsampling patterns $\Omega$ used in this study are equispaced and same for all subjects. However, subsampling pattern $\Omega$ may vary per subject, i.e. indexed by $i$, if random subsampling is used.

**Proposed Self-supervised Training without Fully-Sampled Reference Data**

As discussed previously, acquiring fully sampled data is often difficult or impossible in many scenarios, due to constraints such as organ motion, signal decay or lengthy scan times. Such cases pose an important challenge for the practicality of DL-MRI reconstruction methods that rely on supervised training, since ground truth data is not available for training. To tackle this problem, we propose a self-supervised approach illustrated in **Figure 2**, where the acquired sub-sampled data indices, $\Omega$ from each scan is divided into two sets $\Theta$ and $\Lambda$ as

$$\Omega = \Theta \cup \Lambda. \qquad (9)$$

The set of k-space locations specified by $\Theta$ are used within the network during training in the data consistency units, while the set of k-space points in $\Lambda$ are used to define the loss function. Thus, to

enable training without using fully-sampled data, the following loss function is minimized

$$\min_{\boldsymbol{\theta}} \frac{1}{N} \sum_{i=1}^{N} \mathcal{L}\left(\mathbf{y}_\Lambda^i, \mathbf{E}_\Lambda^i\left(f(\mathbf{y}_\Theta^i, \mathbf{E}_\Theta^i; \boldsymbol{\theta})\right)\right). \tag{10}$$

In other words, the unrolled network output image $f(\mathbf{y}_\Theta^i, \mathbf{E}_\Theta^i; \boldsymbol{\theta})$ which only uses the indices specified by $\Theta$ for data consistency is transformed to k-space using the encoding operator, $\mathbf{E}_\Lambda^i$ specified by the k-space indices in $\Lambda$. Then the loss is calculated in k-space with respect to the acquired k-space data at these locations. In the proposed SSDU approach, $\Theta$ was chosen as $\Omega \backslash \Lambda$. Thus, in our self-supervised training methodology, the unrolled network only sees the acquired k-space data at locations $\Theta = \Omega \backslash \Lambda$ to enforce data consistency. The quality of the final reconstruction, i.e. the network output image, is then checked by mapping to the individual coil k-spaces via $\mathbf{E}_\Lambda^i$, and checking the discrepancy to these acquired measurements at these remaining locations $\Lambda$. Thus, the network is trained to decrease the discrepancy between the network output transformed to all the coil k-spaces and the acquired measurements that it does not see within its unrolled data consistency units. After the network is trained with our proposed self-supervised approach, the reconstruction for unseen test data is performed by using all available measurements at locations $\Omega$.

Our proposed self-supervised approach share similarities with the widely used concept of cross-validation. In machine learning, cross-validation is commonly used to evaluate how accurately a model will perform with robustness to bias and over-fitting issues. Cross-validation is performed by partitioning available data into two sets, one of which is used to train the model and the other for validation, i.e. check whether the trained model generalizes to unseen data. The key difference between our approach and cross-validation is that we perform partitioning per each slice in the dataset, whereas in cross-validation the whole dataset is partitioned only once. The key hyper-parameter for success of cross-validation is the number of folds, which should be well-designed (54). Similarly, in our proposed self-supervised approach, subset selection mechanisms for $\Lambda$ and $\Theta$ are critical, which are thoroughly studied in the next section.

## Methods

### Network and Training Details

The network for solving sub-problems [5] and [6] was unrolled for 10 iterations. The data consistency in the unrolled network was implemented with conjugate gradient method for solving Equation [6], which itself was unrolled for 10 iterations. The neural network for solving the sub-problem [5] was implemented using a convolutional neural network (CNN) based on a ResNet structure, which has shown success in other regression problems (55). This CNN, shown in **Figure 1c**, consisted of a layer of input and output convolution layers, and 15 residual blocks (RB) with skip connections that facilitate information flow during network training. Each RB comprised of two convolutional layers in which the first layer is followed by a rectified linear unit (ReLU) and second layer is followed by a constant multiplication layer, with factor C = 0.1 (55). All layers had a kernel size of 3×3 and 64 channels. This ResNet CNN had a total of 592,129 trainable parameters, which were shared across the unrolled iterations. Coil sensitivity maps were generated from the 24×24 center of k-space using ESPIRiT (56) using a kernel size of 6×6, as well as thresholds of 0.02 and 0.95 for calibration-matrix and eigenvalue decomposition.

A normalized $\ell_1$-$\ell_2$ loss, defined as

$$\mathcal{L}(\mathbf{u},\mathbf{v}) = \frac{\|\mathbf{u}-\mathbf{v}\|_2}{\|\mathbf{u}\|_2} + \frac{\|\mathbf{u}-\mathbf{v}\|_1}{\|\mathbf{u}\|_1}, \qquad (10)$$

was used for both the supervised and the proposed self-supervised training. In the supervised setting, **u** and **v** correspond to the reference ground-truth image/fully-sampled k-space and network output image/network output k-space obtained by transforming network output images to k-space by applying a fully-sampled encoding operator, while for the proposed self-supervised training these correspond to the acquired k-space measurements at locations specified by $\Lambda$ and the k-space corresponding to the network output image at the same locations. For supervised training, k-space loss was used throughout the study as it outperforms the image domain loss used in our preliminary results (57) (**Supporting Information Figure S2**), while also matching our self-supervised framework. Prior to processing, maximum absolute value of the k-space datasets was normalized to 1 in all cases. The networks were trained using the Adam optimizer with a learning rate of $10^{-3}$ unless specified otherwise, by minimizing the corresponding loss function with a batch size of 1 over 100 epochs. All training was performed using Tensorflow in Python, and processed on a workstation with an Intel E5-2640V3 CPU (2.6GHz and 256 GB memory), and an NVIDIA Tesla V100 GPU with 32 GB memory.

**Choice of the Loss Mask**

The proposed SSDU approach divides the acquired sub-sampled data into two disjoint sets $\Theta$ and $\Lambda$. Furthermore, in our implementation, $\Lambda$ is allowed to vary for each different slice in the training database, i.e. they can be indexed as $\{\Lambda_i\}_{i=1}^N$. The subset $\Lambda$ is retrospectively selected from the acquired k-space points, $\Omega$ in order to define the loss function. Hence, unlike the data acquisition process for sampling k-space locations $\Omega$, which is affected by concerns about contrast changes or eddy current artifacts (9), selection of $\Lambda$ is not limited by any physical constraints. This is because $\Lambda$ is selected after data acquisition and amounts to the selection of an index set from all possible acquired k-space locations. Thus, distribution and size of $\Lambda$ were the two hyper-parameters that were studied. For the distribution of $\Lambda$, a uniformly random selection among elements of $\Omega$, as well as a variable density selection based on Gaussian random weighting were investigated. For its size, the ratio $\rho = |\Lambda|/|\Omega|$ was varied among 0.05, 0.1, 0.2, …,0.8, 0.9, where $|\cdot|$ is the cardinality of the index set. A 5-fold cross-validation was also performed on training data for quantitative assessment of the distribution of $\Lambda$, as well as a subset of $\rho$ values among 0.1, 0.2, 0.3, 0.4, 0.5, 0.6.

Additionally, the impact of the overlap between $\Theta$ and $\Lambda$ on the reconstruction performance was also studied. The first scenario considered was the limiting case when $\Omega=\Theta=\Lambda$. Subsequently, we created three different partial overlap scenarios for the best performing $\rho$ value as: 1) The first case, referred to as disjoint sets, in which there is no overlap between $\Theta$ and $\Lambda$ (as originally proposed); 2) The second case, referred to as 50% overlap, where we included 50% of points from $\Lambda$ in $\Theta$ as well. More formally, i.e. $|\Lambda \cap \Theta| / |\Lambda| = 0.5$; 3) Lastly, we have the 100% overlap case where all points in $\Lambda$ is included in $\Theta$ as well (in this case $\Omega = \Theta$, but $\Lambda$ is a subset of $\Omega$).

**Fully-Sampled Knee MRI**

Knee dataset were obtained from the New York University (NYU) fastMRI initiative database (58). Fully sampled raw data were acquired on a clinical 3T system (Magnetom Skyra, Siemens, Erlangen, Germany) with a 15-channel knee coil using 2D turbo spin-echo sequences. The imaging parameters used for the knee data acquisitions are provided in the **Supporting Information Table S1**.

The fully-sampled raw data were under-sampled retrospectively for both training and testing using equispaced sampling patterns provided in the fastMRI database with an acceleration rate (R) = 4 (27,58,59). The center of k-space was fully-sampled with 24 lines of auto-calibrated signal (ACS). The training set consisted of 300 slices from 15 subjects for coronal PD, coronal PDFS, and 10 subjects for sagittal PD, sagittal $T_2$, axial $T_2$. Testing was performed on all slices from 10 different subjects for all knee sequences. Ground truth images for supervised training were generated with a SENSE-1 combination of the fully-sampled data (31,32). The proposed self-supervised approach was compared with supervised DL-MRI trained on fully-sampled dataset and conjugate gradient SENSE (CG-SENSE) (60). Additionally, comparison to a multi-coil compressed sensing reconstruction incorporating coil sensitivities with total generalized variation (TGV) as regularizer (45) was carried out for illustration purposes. However, TGV was not performed on all test datasets since it is computationally expensive, and a comparison between supervised DL-MRI and TGV was already performed in (27). For TGV, the MATLAB implementation provided by authors was utilized (45). We note that TGV and CG-SENSE approaches are shown only for comparison purposes with more traditional methods, and are not considered as competitive baseline images, consistent with previously reported results in the literature (27).

**Prospectively Accelerated Brain MRI**

Brain imaging was performed on 19 healthy subjects at a 3T Siemens Magnetom Prisma (Siemens Healthcare, Erlangen, Germany) system using a 32-channel receiver head coil-array. The imaging protocols were approved by the local institutional review board, and written informed consent was obtained from all participants before each examination for this HIPAA-compliant study. Data acquisition was performed using a standard Siemens 3D-MPRAGE sequence with the following parameters: FOV = 224×224×157 mm$^3$, resolution = 0.7×0.7×0.7 mm$^3$, TR/TE = 2400 ms/2.2 ms, inversion time = 1000 ms, flip angle = 8°, band-width = 210 Hz/pixel, 3D matrix size = 320×320×224, prospective acceleration R = 2 (equispaced in $k_y$), ACS lines = 32, acquisition orientation = sagittal. The k-space data was inverse Fourier transformed along the read-out (foot-head) direction, and these axial slices were processed individually. The prospectively undersampled brain datasets were further retrospectively undersampled to R = 4, 6, 8 using a sheared equispaced $k_y$-$k_z$ undersampling pattern (61), with a 32×32 ACS region in the $k_y$-$k_z$ plane. Sampling masks are provided in **Supporting Information Figure S3**. We note that while in principle prospectively sub-sampled data can be acquired at all these different rates, we chose to utilize further retrospective sub-sampling of prospectively accelerated data since our focus is on the reconstruction quality and this approach avoids confounding factors between different scans, such as subject motion or variations from $T_1$ recovery. We also note that when the self-supervised approach was used at one of these higher acceleration rates, it only had access to the k-space data corresponding to that acceleration rate, both during training and testing. The learning rate for training was set to $5 \cdot 10^{-4}$. The training set consisted of 300 slices from 10 subjects, formed by taking the central 30 slices from each subject. Testing was performed on all slices from 9 different subjects.

The proposed self-supervised DL-MRI results were compared to CG-SENSE method. We note that a comparison to supervised DL-MRI was not possible in this setting, since there was no fully-sampled ground truth data.

**Image Evaluation**

Experimental results were quantitatively evaluated using normalized mean square error (NMSE) and structural similarity index (SSIM). Additionally, qualitative assessment of the image quality was performed by an experienced radiologist. For knee MRI, the proposed self-supervised DL-MRI approach was compared to ground truth fully-sampled images, supervised DL-MRI trained on fully-sampled data and CG-SENSE at the same acceleration R = 4. As noted earlier, TGV was not included in the comparison due to its computational complexity and availability of a previous study comparing supervised DL-MRI and TGV (27). For brain MRI, proposed self-supervised DL-MRI reconstructions at acceleration R = 4, 6 and 8 were compared with CG-SENSE approach at the acquisition acceleration R = 2. The reader was blinded to the reconstruction method, except for the knowledge of the reference image in knee MRI datasets. The order in which the methods were shown was also randomized. There were differences between the sequences used for the fastMRI database and our institutional sequences, thus this knowledge allowed the radiologist to assess the baseline image quality. All five knee MRI weightings and brain dataset were evaluated on a 4-point ordinal scale, adopted from (27) for blurring (1: no blurring, 2: mild blurring, 3: moderate blurring, 4: severe blurring), SNR (1: excellent, 2: good, 3: fair, 4: poor), aliasing artifacts(1: none, 2:mild, 3: moderate, 4: severe) and overall image quality (1: excellent, 2: good, 3: fair, 4: poor). Wilcoxon signed-rank test was used to evaluate the scores with a significance level of $P < 0.05$.

# Results

## Choice of the Loss Mask

**Figure 3** depicts the self-supervised network training using varying subsets across slices by uniformly random and variable-density Gaussian selection of $\Lambda \subset \Omega$ for $\rho = 0.1$. Uniformly random selection of $\Lambda$ suffers from visible residual artifacts, marked by red arrows. These artifacts are further suppressed in the Gaussian-based approach and difference images align with these observations. The quantitative assessment from 5-fold cross-validation are consistent with these qualitative assessments. The median and interquartile range of SSIM values were 0.9380 [0.9197, 0.9527], 0.9457 [0.9293, 0.9575], and NMSE values were 0.0021 [0.0016, 0.0027], 0.0019 [0.0015, 0.0023] using uniform random selection and Gaussian selection, respectively. **Supporting Information Figure S4** shows additional reconstructions for uniform random and Gaussian selection for different $\rho$ values, which further highlights that Gaussian selection consistently outperforms uniform random selection across different $\rho$ values. Thus, a variable-density Gaussian selection was used for $\Lambda$ for the remainder of the study.

**Figure 4** shows the impact of network training with varying $\rho \in 0.05, 0.1, 0.2, \ldots, 0.8, 0.9$ using variable-density Gaussian selection. Red arrows show visible residual artifacts for low $\rho$ values of 0.05, 0.1, 0.2. As cardinality of $\Lambda$ increases towards $\rho = 0.4$, residual artifacts decrease. At $\rho = 0.4$, visible artifacts seen at lower $\rho$ values are further suppressed. Residual artifacts start to reappear starting from $\rho = 0.5$, and these artifacts become more pronounced as $\rho$ increases. The quantitative assessment from 5-fold cross-validation aligns with these qualitative assessments. The median and interquartile range of SSIM values were 0.9457 [0.9293, 0.9575], 0.9477 [0.9323, 0.9591], 0.9488

[0.9328, 0.9603], 0.9507 [0.9352, 0.9614], 0.9450 [0.9297, 0.9569], 0.9391 [0.9225, 0.9524], and NMSE values were 0.0019 [0.0015, 0.0023], 0.0018 [0.0013, 0.0023], 0.0018 [0.0014, 0.0022], 0.0017 [0.0013, 0.0021], 0.0020 [0.0015, 0.0024], 0.0022 [0.0016, 0.0028] using Gaussian selection for $\rho \in 0.1, 0.2, 0.3, 0.4, 0.5, 0.6$, respectively. Hence, $\rho = 0.4$ was used for the remainder of the study.

**Figure 5** shows the impact of different degrees of overlap between $\Lambda$ and $\Theta$ for $\rho = |\Lambda|/|\Omega| = 0.4$, as well as the limiting case that uses all available data for both data consistency and loss (i.e. $\Omega=\Theta=\Lambda$). For the limiting case with $\Omega=\Theta=\Lambda$, the reconstruction results suffer from residual noise amplification. On the other hand, when $\Lambda$ and $\Theta$ were disjoint as proposed, such noise amplifications are significantly suppressed. Quantitative SSIM and NMSE evaluation of these methods over the dataset are presented in **Supporting Information Table S2**, indicating that for different rates of overlap between $\Lambda$ and $\Theta$ with $\rho = 0.4$, the performance degrades as the amount of overlap increases. Thus disjoint sets were used for the remainder of the study.

**Knee MRI**

**Figure 6** demonstrates the reconstruction results of coronal PD images using CG-SENSE, TGV, supervised DL-MRI and proposed self-supervised DL-MRI approach along with the ground truth reference, as well as difference images with respect to this reference. CG-SENSE and TGV suffer from visible residual artifacts, marked by red arrows, with the latter having fewer artifacts. The proposed self-supervised and supervised DL-MRI approaches successfully remove the residual artifacts, while achieving similar qualitative and quantitative performance. Quantitative metrics and difference images displayed in the figure are in agreement with these observations. **Supporting Information Figure S5** shows the training loss curves for both approaches where loss decreases over epochs in a similar trend.

The same trends were observed for coronal PD-FS as depicted in **Figure 7**. Both proposed and supervised DL-MRI approaches show similar performance, while improving the suppression of residual artifacts that are visible in CG-SENSE and TGV methods. Quantitative evaluation and the residual artifacts apparent in the difference images also highlight these observations. **Supporting Information Figure S6** show reconstruction results for axial $T_2$, sagittal $T_2$ and sagittal-PD weighted knee dataset which align with observation from coronal weighted knee datasets.

**Figure 8** shows a box-plot displaying the median and interquartile range ($25^{th}$-$75^{th}$ percentile) of the quantitative metrics, SSIM and NMSE, across all test datasets for each knee sequence. In all sequences, supervised and self-supervised DL-MRI approaches achieve similar quantitative performance for both SSIM and NMSE, while significantly outperforming the CG-SENSE approach. We note again that TGV was not included in these comparisons, as it is computationally expensive, and a comparison between supervised DL-MRI and TGV was already performed in (27).

**Prospectively Accelerated Brain MRI**

**Figure 9** depicts a sagittal slice of the 3D MPRAGE dataset at acquisition acceleration R = 2 and further retrospective acceleration R = 4, 6 and 8 reconstructed with CG-SENSE, as well as R = 4, 6 and 8 reconstructed with the proposed self-supervised DL-MRI on a representative test subject, following reformatting to the original acquisition (sagittal) plane. CG-SENSE suffers from significant noise amplification at higher acceleration rates. Self-supervised DL-MRI successfully performs reconstruction at these higher acceleration rates, while achieving lower noise level and similar overall image quality with CG-SENSE at R = 2. Results from another subject are depicted in **Supporting Information Figure S7** and shows similar trends. TGV was not applied due to the high computational runtime across all axial slices, and supervised DL-MRI cannot be applied in this setting due to the lack of fully-sampled references.

**Image Evaluation Scores**

**Figure 10** summarizes the results of the reader study for knee and brain datasets. For knee datasets, both supervised and self-supervised DL-MRI approaches get comparable scores to the reference image in terms of SNR, blurring, aliasing artifacts and overall image quality. There was no statistical difference between reference and DL-MRI approaches in terms of the evaluation criterions for all knee sequences, except for blurring between reference and DL-MRI approaches in coronal PD-FS. CG-SENSE was significantly outperformed by both DL-MRI approaches, while showing statistically significant differences to the reference and both DL-MRI approaches for all knee sequences, except in blurring for coronal PD and PD-FS sequences. More comprehensive bar plots of the average scores including CG-SENSE and supervised training with image domain loss as in Equation [7] are presented in **Supporting Information Figure S8**.

For the 3D MPRAGE dataset, DL-MRI reconstructions trained using the proposed self-supervised approach at acceleration rates 4, 6 and 8 show similar statistical properties in terms of SNR and blurring with CG-SENSE at acquisition R = 2. However, in terms of aliasing artifacts and overall image quality, proposed self-supervised approach at all three acceleration rates (R = 4, 6 and 8) outperform CG-SENSE at R = 2. In terms of aliasing artifacts, proposed self-supervised approach for rates 4 and 6 show similar statistical behavior with each other, while significantly improving upon self-supervised DL-MRI at R = 8 and CG-SENSE at R = 2, which perform statistically similar among themselves. Proposed self-supervised approach at R = 4 shows the best overall image quality and shows statistically significant differences with self-supervision at R = 6, 8 and CG-SENSE at R = 2. As expected, the overall image quality decreases with higher acceleration rates using the proposed self-supervised DL-MRI approach, although these techniques still outperform CG-SENSE at R = 2.

## Discussion

In this study, we developed a framework for self-supervised training of physics based DL-MRI reconstruction without fully sampled data. The proposed approach split the acquired under-sampled k-space indices into two disjoint sets $\Theta$ and $\Lambda$, where the former was used across the unrolled network to enforce data consistency, while the latter was used to define the loss function for the training. The results on retrospectively under-sampled knee datasets showed that our SSDU approach achieves comparable results with a supervised DL-MRI approach using the same neural network architecture, while outperforming conventional CG-SENSE and TGV approaches. Results on prospectively under-sampled brain datasets, for which supervised learning methods cannot be applied due to unavailability of fully-sampled data, further confirmed the effectiveness of the proposed self-supervised training approach for DL-MRI reconstruction. These reconstructions at higher acceleration rates of 4, 6 and 8, visually outperformed CG-SENSE at R = 2 according to the reader study.

Most DL-MRI approaches use supervised learning for network training in order to provide improved accelerated MRI reconstruction (28,29,32,33,59). However, acquiring fully-sampled data is challenging in many practical scenarios of interest. These may be due to constraints on timing, physiological constraints, signal decay or long scan times (38-42). As an example, the fully-sampled acquisition for the 3D MPRAGE sequence with the resolution used in this study would be more than 15 minutes (41), which is impractical for large studies and may lead to patient discomfort. Furthermore, such long scan times increase susceptibility to motion artifacts, which would be more pronounced at these high resolutions. To further highlight the need for training data, we have also performed experiments on prospectively sub-sampled snapshot cardiac MRI, where it is infeasible to collect the ground truth data. Results from these experiments are shown in **Supporting Information Figure S9**, showing the applicability of our method in this setting as well. Thus, being able to train DL-MRI reconstruction methods without fully-sampled data is imperative to broaden their application to settings in which such data is challenging to acquire, where supervised training are no longer practical. Furthermore, this may also facilitate the integration of DL-MRI methods to many clinical scans that readily include a form of accelerated imaging, most commonly in the form of parallel imaging, by enabling the use of prospectively undersampled raw k-space data for training.

Given the importance of training without fully sampled data, there have been several works which have tried to tackle this issue. For purely data-driven de-aliasing of single-coil data using image domain to image domain mapping without the encoding operator, a self-supervised approach has been proposed (62) using a mixture of measurement and k-space losses. Unlike our approach, it uses all available data for training and loss, i.e. identical sets. As a result, the reconstructions suffer from visible noise amplifications which also align with our observation about usage of identical sets in **Figure 5**. An alternative approach, which assumes the same data is acquired with two separate acquisitions using different undersampling patterns was also proposed (63,64) extending on the Noise2Noise denoising framework (65). In the same image-domain reconstruction setting, a self-supervised learning scheme using cycleGANs with optimal transport cost minimization was proposed (66), although initial results exhibit blurring artifacts. Although purely data-driven image domain methods have been used for DL-MRI reconstruction, physics-guided DL-MRI techniques are more desirable as they offer a degree of interpretability by incorporating domain knowledge on the MRI encoding mechanism (20,27,28,30,31,33). In this physics-guided setting, an unpaired learning approach using Wasserstein GANs was proposed (67), but this procedure still assumes the presence of high-quality images albeit not requiring pairwise matching with undersampled data. Another approach uses the so-called unsupervised basis pursuit (68,69), where the unrolled network consists of regularizer units followed by several consecutive DC units. This approach uses the current output of the DC unit as the training label, and iteratively updates both network parameters and this training label, in a method reminiscent of semi-supervised training. This method was investigated with random undersampling patterns, where intermediate outputs tend to suffer from noise amplification but without significant residual artifacts. In this setting, this approach was able to reduce noise further, even though noise amplification was observed when compared to supervised training (68,69). However, this method was not investigated for equispaced undersampling, as is the focus of this study, where intermediate DC outputs are both noisy and likely to have residual aliasing artifacts. Thus, the utility of this method in equispaced undersampling is unclear and warrants further investigation. In contrast, our SSDU approach uses physics-guided DL-MRI reconstruction, while not making any explicit assumptions about the final output in image space. In particular, we do not enforce the output of our network to align with a generative model or consider intermediate estimates as reference output for training. The training in SSDU only considers the acquired k-space data to evaluate the reconstruction quality, in effect using a physics-guided self-supervision approach. Furthermore, SSDU works for both equispaced undersampling patterns, as is the focus of the study, and random undersampling patterns (results not shown). Note the former was considered to be more challenging for physics-guided DL-MRI reconstruction in previous studies, as networks trained with equispaced sampling were shown to generalize well to random sampling, but not the vice versa (27,70).

Our training method is also reminiscent of the broader and fundamental concept of cross-validation in machine learning and statistics (71). When testing generalizability, the training database is partitioned into two sets of complementary datasets, one which is used for training the model (often called training set), and the other used to assess the performance in unseen data (often called validation/testing set). In our approach, we do a similar partitioning of the acquired data to two sets we denoted $\Theta$ and $\Lambda$. The main difference to typical cross-validation is that our partitioning is done for each subject in the training set from the database. But the intuition for partitioning within the network is similar, as the unrolled network only sees $\Theta$ for data consistency during training, while $\Lambda$ is only used to establish the network loss. Indeed, as our experiments in **Figure 5** show that when $\Theta$ and $\Lambda$ are taken to be the same as $\Omega$, such training leads to poor image quality with insufficient removal of aliasing artifacts and noise amplification, as the DC unit operating on the full $\Omega$, inherently matches well with the acquired data at these locations.

Selection of the loss mask, $\Lambda$ plays an important role in the performance of the proposed self-supervised training. One major design advantage is that since it only exists in post-processing, it can be chosen freely among all the acquired measurements retrospectively, without physical constraints that are imposed during acquisition. Thus even though 40% of the acquired indices in Ω were included in Λ, this is not the equivalent to training with an ~8-fold accelerated acquisition, especially for the 2D setting, since the points in Λ do not need to constitute fully-sampled readout encoding lines along $k_x$. This point is further illustrated in **Supporting Information Figure S10**, in the context of supervised training. This advantage is not as clear in the training for the 3D brain dataset in this study, since the data had to be inverse Fourier transformed along the foot-head readout direction and axial slices had to be processed due to memory issues in the GPUs. In this case, the sheared equispaced $k_y$-$k_z$ undersampling pattern readily do not include any lines, thus the selection of Λ, may affect the DC units more substantially than in the 2D knee MRI experiments. Accordingly, the self-supervised approach is expected to show more gains and better reconstruction quality at higher acceleration rates for 3D imaging if 3D neural networks can be used. Thus memory-efficient 3D neural network designs (72) may warrant further investigation, although it is beyond the scope of the current study.

The data reduction arising from data splitting between Θ and Λ poses more challenges for training and reconstruction at higher acceleration rates, even for 2D acquisitions. This was further investigated to check how the performance of self-supervised and supervised training would change at higher acceleration rates when all the training parameters and datasets are the same as described earlier. The results shown in **Supporting Information Figure S11** indicate that both training methods perform similarly at R = 4 and 6 for knee MRI. However, at R = 8, where the supervised training is able to suppress artifacts albeit at the cost of blurring artifacts, the self-supervised approach starts suffering from additional residual aliasing artifacts. Thus, at higher acceleration rates, where reconstructions from the supervised training can operate without aliasing artifacts but with quality degradation, the self-supervised approach faces additional challenges including residual aliasing, due to the scarcity of data, especially after the splitting to two sets. The problem of data scarcity has been addressed by several important transfer learning methods when using supervised training with fully-sampled datasets (73,74). These approaches pre-train neural networks on fully-sampled large datasets and then fine-tune them on smaller datasets of interest. In such cases, if the smaller dataset of interest is additionally not fully-sampled, then the proposed self-supervised approach may be combined synergistically with transfer learning to tackle this challenging issue of both data scarcity and not having fully-sampled data, though this was beyond the scope of this study.

All experiments in this study were based on Cartesian acquisitions. The proposed self-supervised approach can be extended to non-Cartesian acquisitions. In non-Cartesian acquisitions such as radial or spiral acquisitions, one can choose the subsets for training and loss mask from the acquired radial spokes and spirals, similar to Cartesian acquisitions used in this study, since this amounts to selecting a subset of individual k-space points on the spokes or spirals. We also note that for non-Cartesian acquisitions, the encoding operator also contains the gridding/de-gridding operation to account for non-uniform Fourier transforms. These extensions were not investigated, as it was beyond the scope of the current work.

In this study, we compared uniformly random selection with a variable-density approach based on Gaussian weighting for selecting $\Lambda$. In our experiments, the latter selection was favored as it statistically outperformed and visibly improved upon the former. A self-supervised mask selection during the network training may further remove these hyper-parameters and potentially lead to further improvements in reconstruction. However, this is a difficult problem, which warrants further investigation, beyond the scope of the current study. Using different distributions for selecting a number of distinct $\Theta$ and $\Lambda$ pairs per subject may further improve performance, but currently these distributions would need to be empirically chosen. Due to the ad-hoc nature of such a process and the wide range of available distributions, this was not explored in detail, but this idea also warrants more investigation in the context of self-supervised mask selection in future works. We also investigated the reconstruction performance using the same sets, $\Theta$ and $\Lambda$, across all training slices versus letting these vary across slices as $\Theta_i$ and $\Lambda_i$, as proposed. Although one can choose these sets to be same for all slices, such an approach bears the risk of a sub-optimal loss mask being used for all slices. Hence, having different sets for each slice in the dataset may provide additional robustness. **Supporting Information Figure S12** shows that having different loss and training sets for each slice shows slight improvement over using the same sets across all the training dataset. Finally, a heuristic choice was made to keep 4×4 central k-space lines in the $\Theta$ set, as the DC units did not work well without these high-energy components. In our experience, use of larger (8×8 or 16×16) or smaller (2×2) regions deteriorated the overall performance.

The same residual network structure for regularizer and unrolled conjugate gradient for data consistency units were used throughout the study. However, our approach is not restricted to these network and DC unit choices. Alternative approaches, such as a DenseNet, U-Net or variational neural network as a regularizer CNN (27,75,76), or gradient descent for the DC unit are also possible (27,33). However, these were not explored, since such network optimization was not the focus of our study. Instead we fixed one architecture, and used this for both supervised and self-supervised training. In this study, we also shared the regularizer CNN parameters across the unrolled network, similar to (28,33), in order to enable training with a smaller training dataset. However, it is possible to use different parameters for each unrolled regularizer unit, as in (27,31), at the cost of a higher number of trainable parameters. A comparison between supervised training with shared and non-shared parameters in the unrolled network is provided in **Supporting Information Figure S13**. The results indicate that the two approaches perform similarly in terms of qualitative and quantitative assessments.

Selection of proper loss functions also play a vital role for network training. The $\ell_2$ loss is a frequently used metric in DL-MRI with promising results (20,28), but it is sensitive to outliers. On the other hand, $\ell_1$ loss is more robust to outliers. Hence, we used a normalized $\ell_1$-$\ell_2$ loss to take advantage of the superior properties of each loss while minimizing their disadvantages (53). Other choices of losses such as discriminative losses have also been popular for supervised training of DL-MRI methods (33,77). There have also been works to incorporate the conventional loss functions such as $\ell_1$ or $\ell_2$ into adversarial losses (25,78-80). To the best of our knowledge, there are no works that use an adversarial loss in k-space, but such an extension may benefit the reconstruction quality when using the proposed self-supervision approach.

## Conclusion

The proposed training framework allows training of physics-guided DL-MRI reconstruction without requiring fully-sampled data, while performing similar to conventional supervised DL-MRI approaches.

## Acknowledgements

Knee MRI data were obtained from the NYU fastMRI initiative database (58). NYU fastMRI database was acquired with the relevant institutional review board approvals as detailed in (58). NYU fastMRI investigators provided data but did not participate in analysis or writing of this report. A listing of NYU fastMRI investigators, subject to updates, can be found at fastmri.med.nyu.edu.

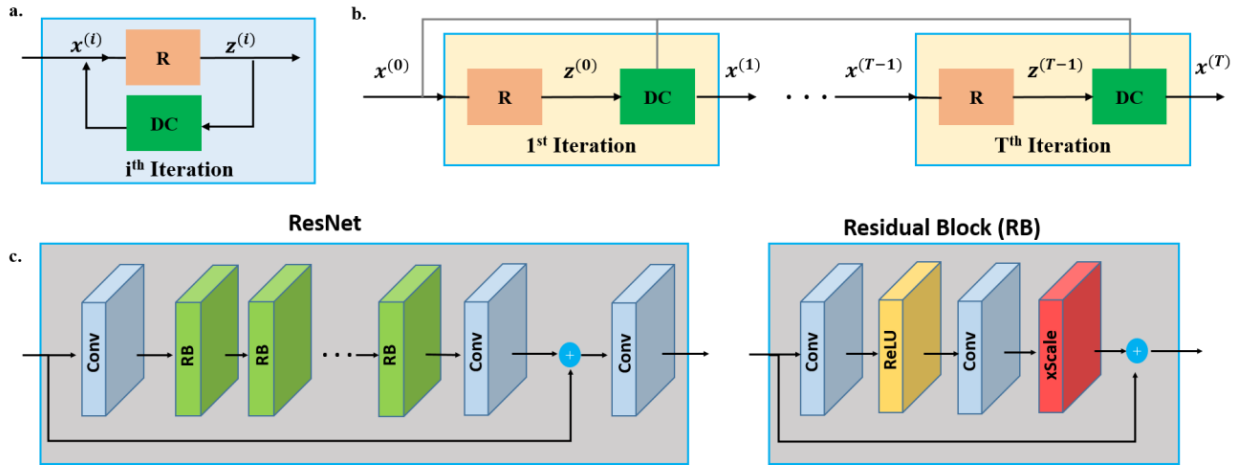

**Figure 1.** a) Depiction of a conventional iterative optimization algorithm for solving regularized inverse reconstruction problems. These algorithms alternate between regularization (R) and data consistency (DC). b) For neural networks, this iterative algorithm is unrolled for $T$ steps, leading to a feed-forward structure alternating between R and DC units, where R is implemented by means of a neural network. c) The ResNet architecture (49) used as regularizer (R) in this study consists of 15 residual blocks (RB), each of which contains two convolution layers with the first one followed by a ReLU and the second one followed by a constant multiplication layer.

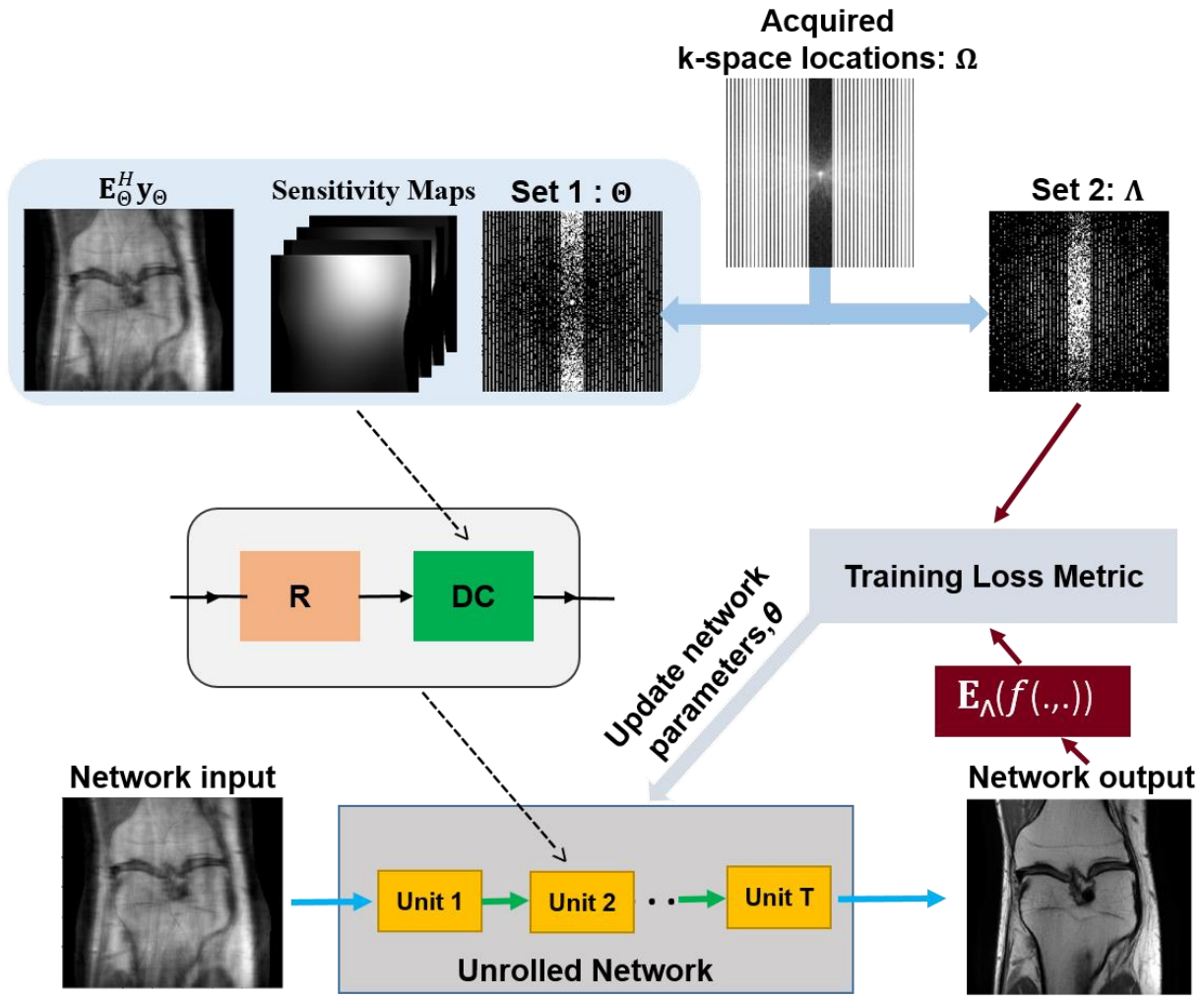

**Figure 2.** The self-supervised learning scheme to train physics-guided deep learning without fully-sampled data. The acquired sub-sampled k-space measurements, $\Omega$, are split into two disjoint sets, $\Theta$ and $\Lambda$. The first set of indices, $\Theta$, is used in the data consistency unit of the unrolled network, while the latter set, $\Lambda$ is used to define the loss function for training. During training, the output of the network is transformed to k-space, and the available subset of measurements at $\Lambda$ are compared with the corresponding reconstructed k-space values. Based on this training loss, the network parameters are subsequently updated.

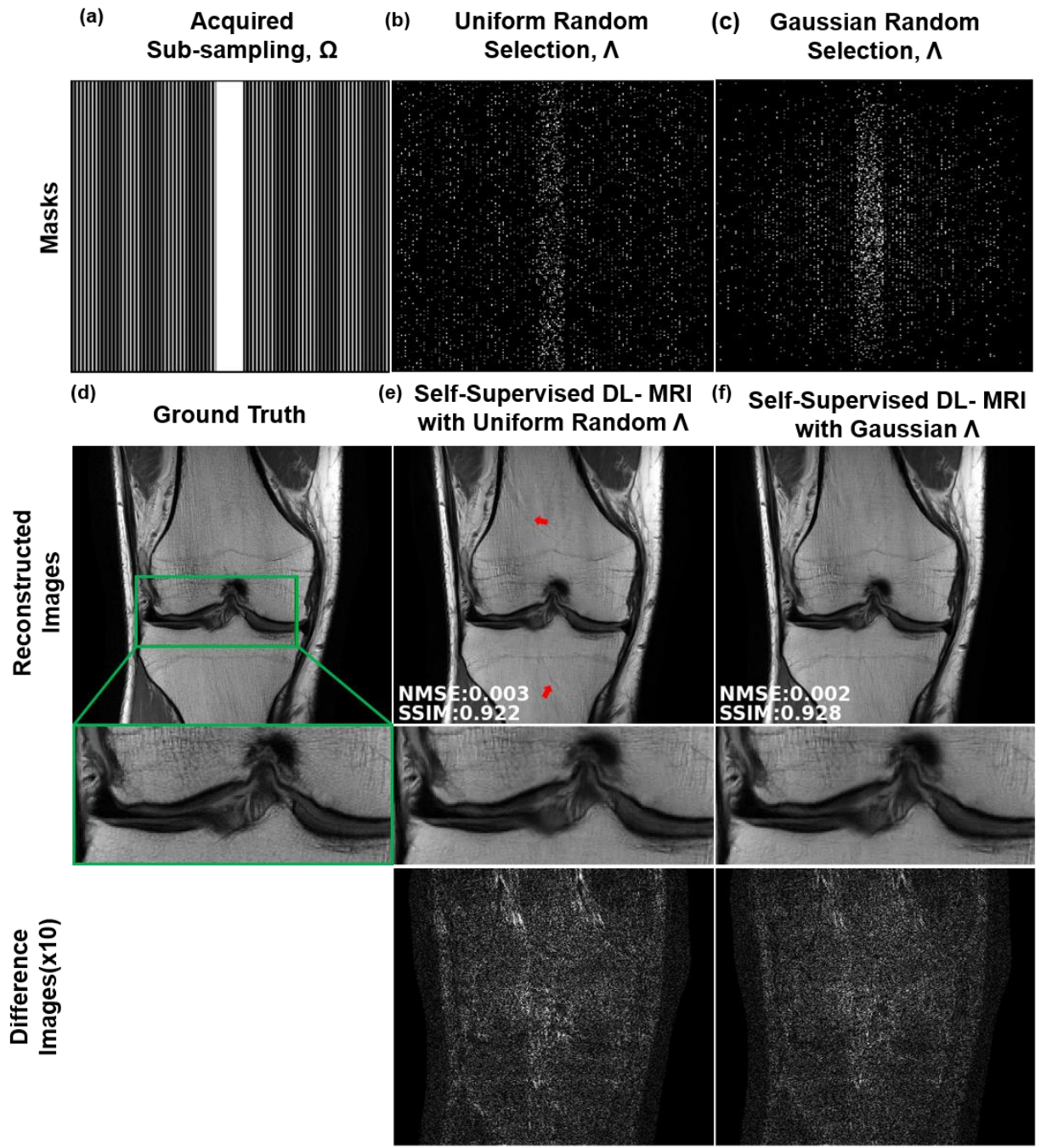

**Figure 3.** a) Acquired sub-sampling pattern, Ω; b) Example uniform random and c) variable-density Gaussian random selection for subset Λ (allowed to differ for each slice in the training dataset) that is used to define the training loss; d) Ground-truth reference data; e) and f) Self-supervised DL-MRI reconstruction and corresponding difference images with loss masks Λ as in

b) and c), respectively. Red arrows mark residual artifacts in uniform random selection. These artifacts are further suppressed in the Gaussian random selection, which is used for the remainder of the study.

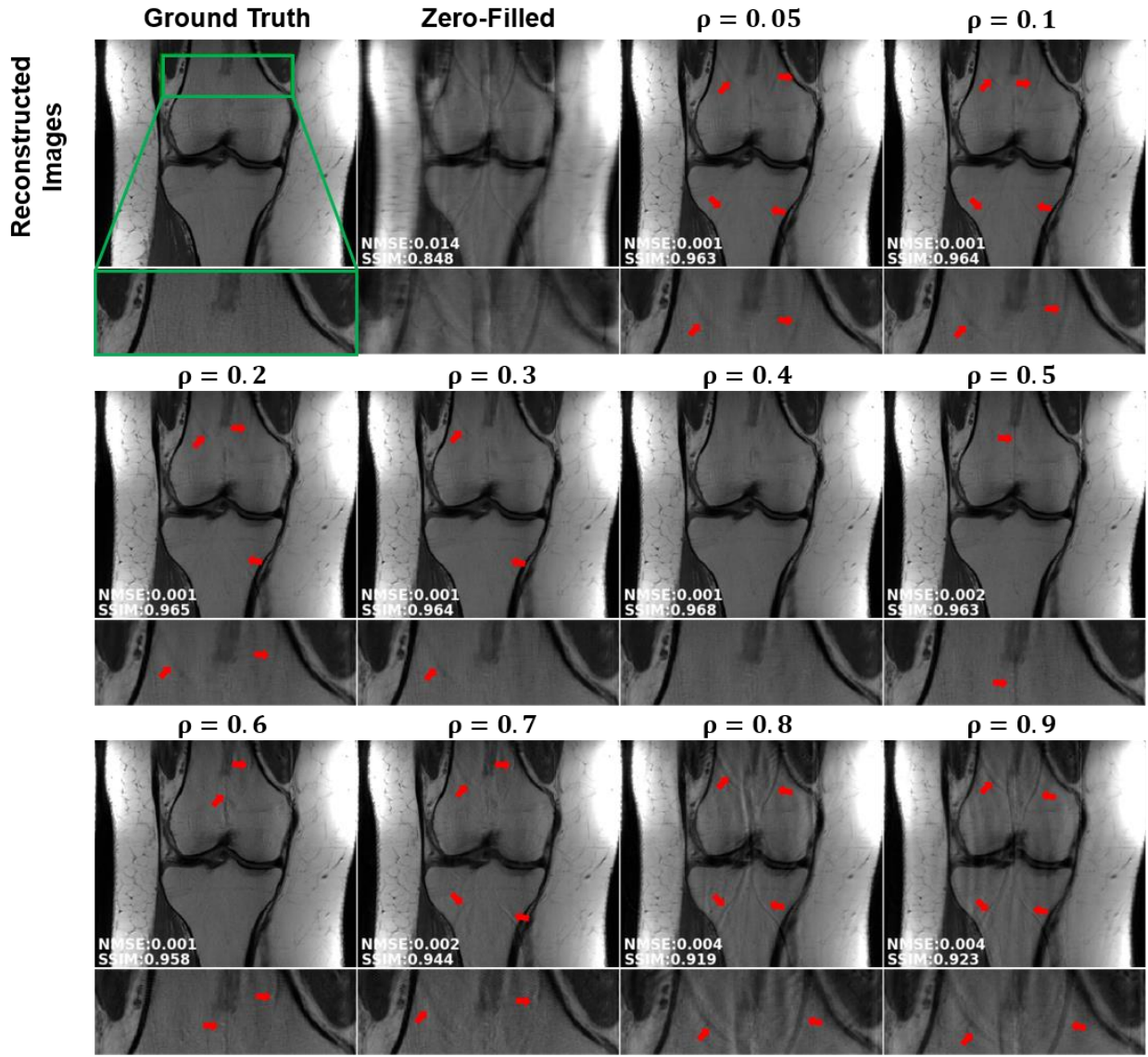

**Figure 4.** A representative test slice depicting the reconstruction results for different ratios of $\rho = |\Lambda|/|\Omega|$. $\Lambda$ is used only for defining loss function, while $\Theta = \Omega \backslash \Lambda$ is only used within data consistency units. Red arrows mark visible residual artifacts for $\rho \leq 0.3$ and $\rho \geq 0.5$. These artifacts are suppressed at $\rho = 0.4$, which is used for the remainder of the study.

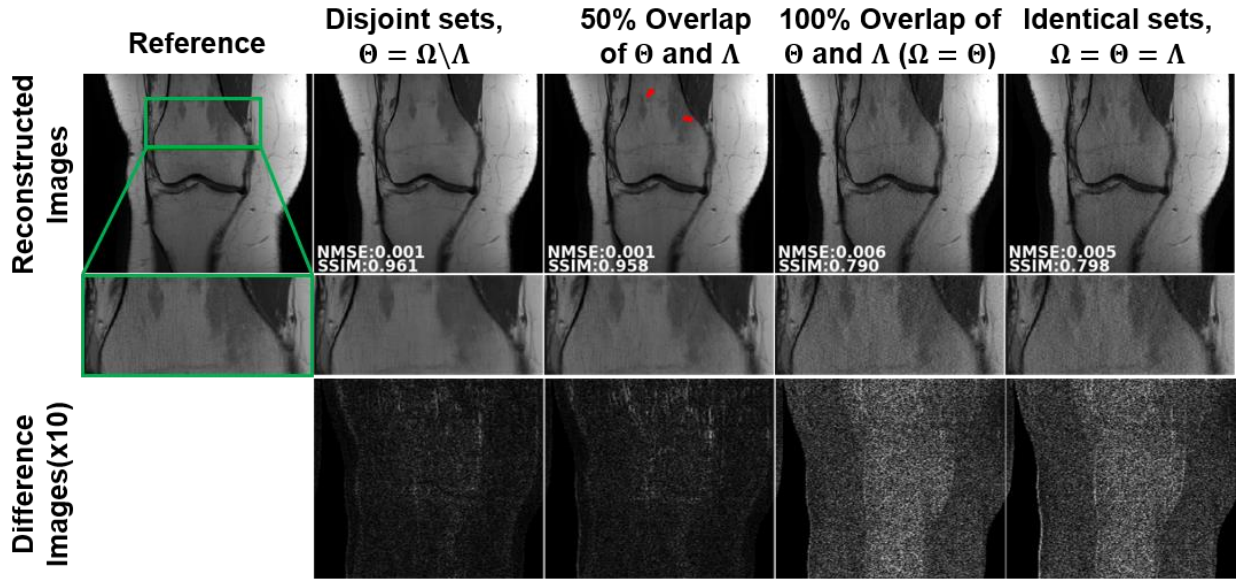

**Figure 5.** Reconstruction results for different degrees of overlap between Λ and Θ, i.e. |Λ∩Θ|/|Λ|, for ρ = |Λ|/|Ω| = 0.4, as well as the limiting case that uses all available data for both data consistency and loss (i.e. Ω=Θ=Λ). For the limiting case with Ω=Θ=Λ, the reconstruction suffers from noise amplification, which is significantly suppressed for the proposed disjoint Λ and Θ. The performance of the self-supervised approach degrades as the amount of overlap increases.

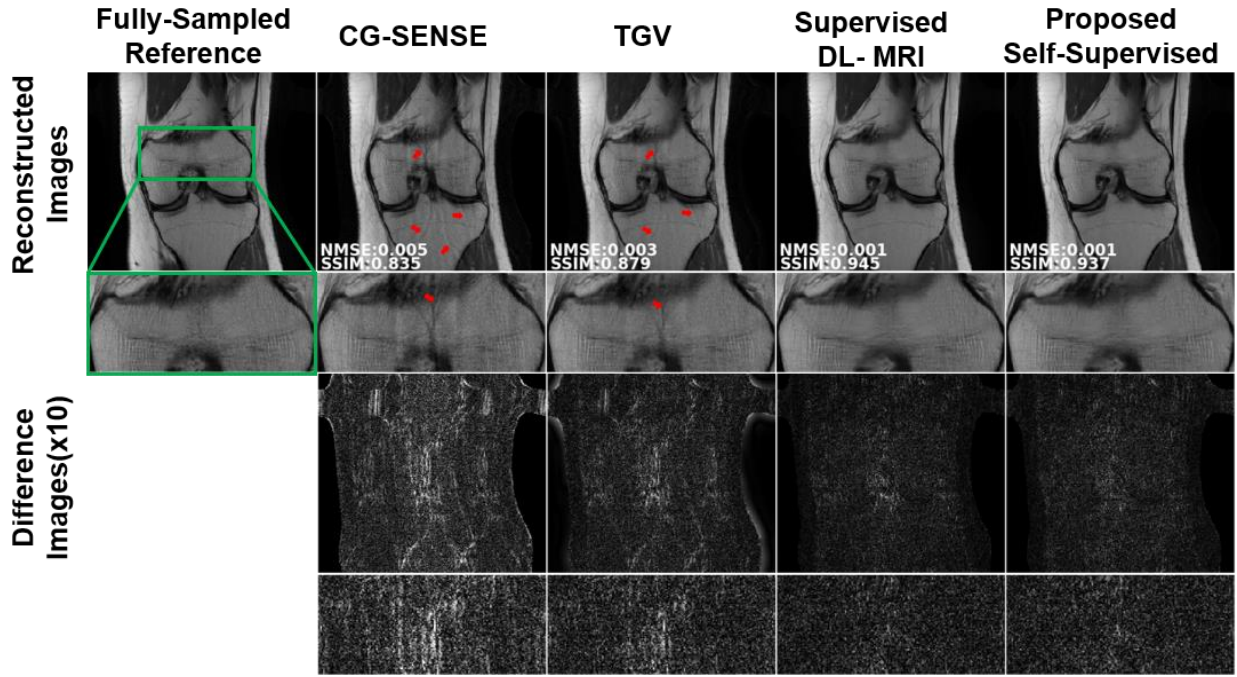

**Figure 6.** A representative test slice from fastMRI coronal PD knee MRI dataset depicting the reconstruction results for proposed self-supervised DL-MRI, supervised DL-MRI, CG-SENSE and TGV approaches for retrospective equispaced undersampling R = 4. Zoomed views and error images show the residual artifacts observed in CG-SENSE and TGV approaches. Both self-supervised and supervised DL-MRI approaches successfully suppress these artifacts, while showing similar quantitative performance.

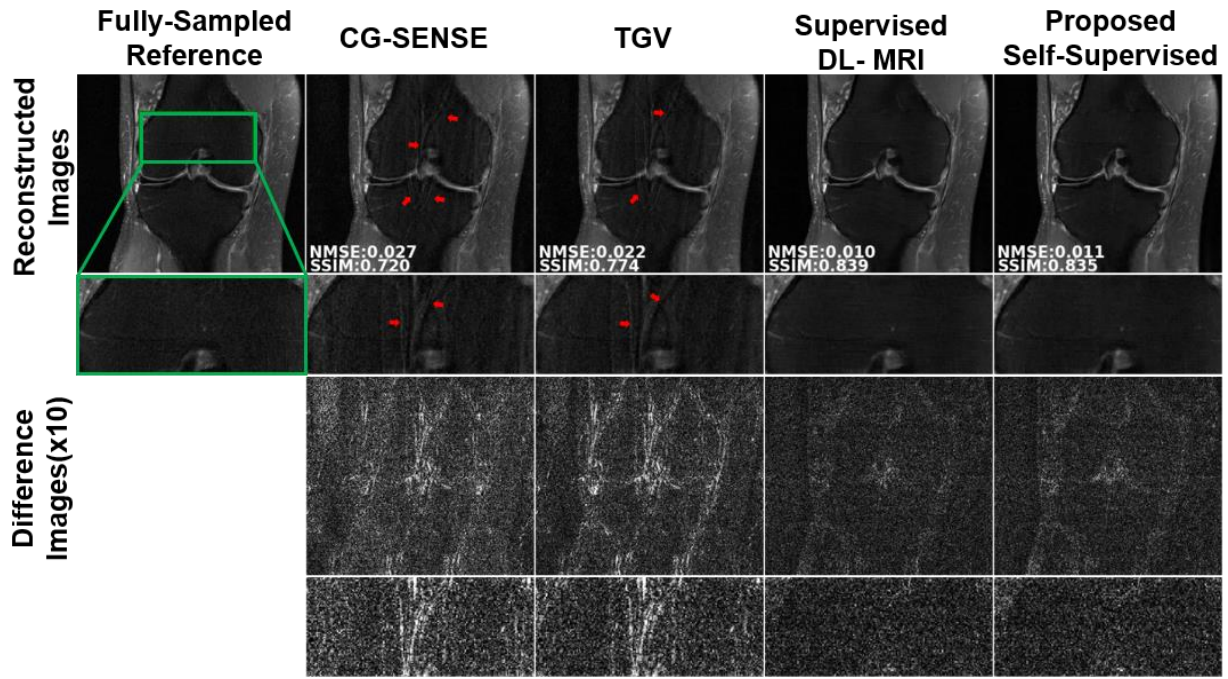

**Figure 7.** A reconstructed test slice showing reconstruction results from fastMRI coronal PD-FS datasets for retrospective equispaced undersampling R = 4. Red arrows indicate visible artifacts, especially apparent in the zoom views and error images for CG-SENSE and TGV techniques. Proposed self-supervised and supervised DL-MRI eliminate these artifacts, while showing similar quantitative and qualitative performance.

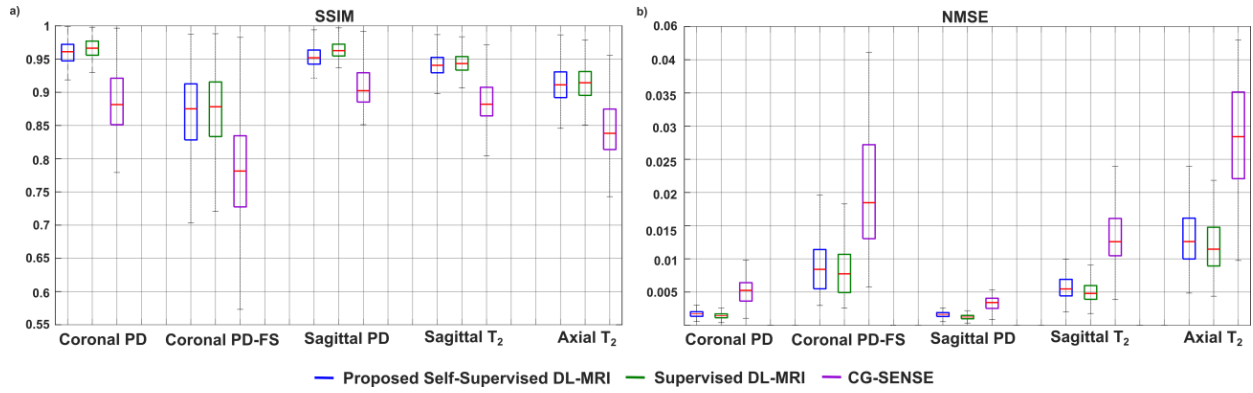

**Figure 8.** Boxplots showing the median and interquartile range (25th-75th percentile) of the quantitative metrics, (a) structural similarity index and (b) normalized mean squared error (NMSE) for all five knee MRI sequences. Both proposed self-supervised and supervised DL-MRI significantly outperform CG-SENSE in terms of SSIM and NMSE for all knee sequences, while showing similar quantitative performance.

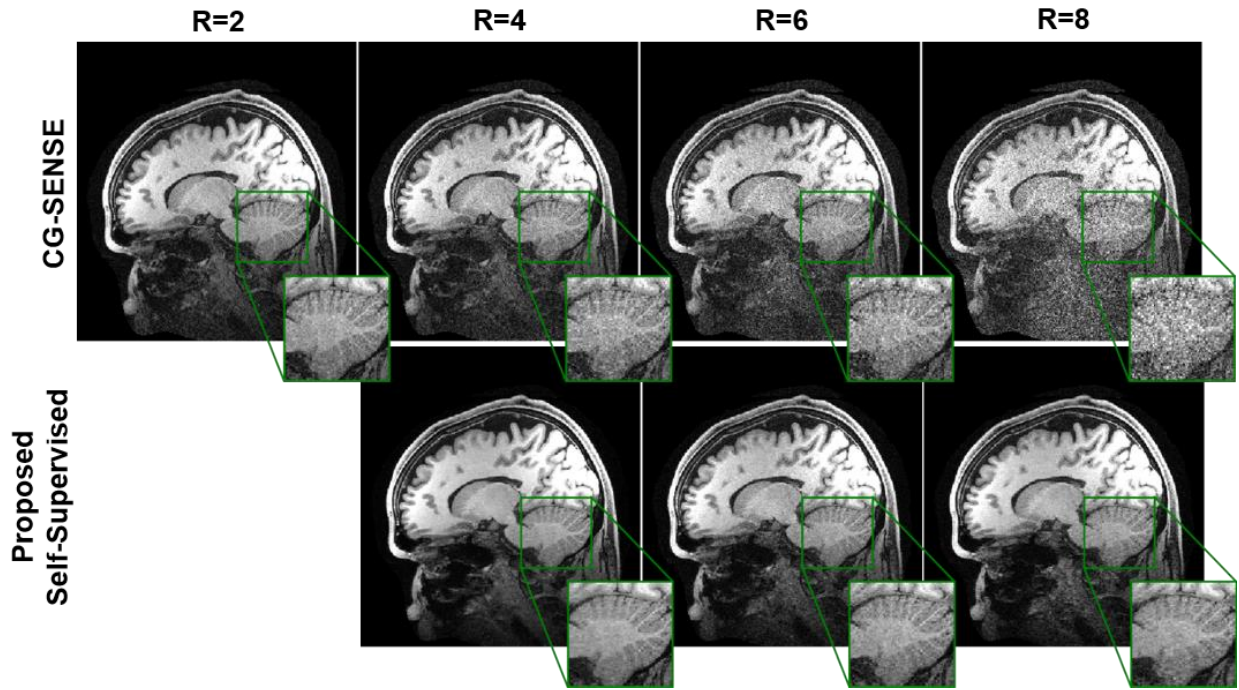

**Figure 9.** Reconstruction results from prospectively 2-fold equispaced undersampled brain MRI. CG-SENSE and the proposed self-supervised approach are applied at further retrospective acceleration rates of 4, 6 and 8 with equispaced sheared $k_y$-$k_z$ undersampling patterns, while CG-SENSE is also used at the acquisition rate of 2. CG-SENSE suffers from visibly higher noise amplification at high acceleration rates. The proposed approach successfully reconstructs brain MRI at these higher rates, achieving similar image quality to CG-SENSE at R = 2. Note the supervised DL-MRI cannot be applied here due to the lack of fully-sampled ground truth data for training.

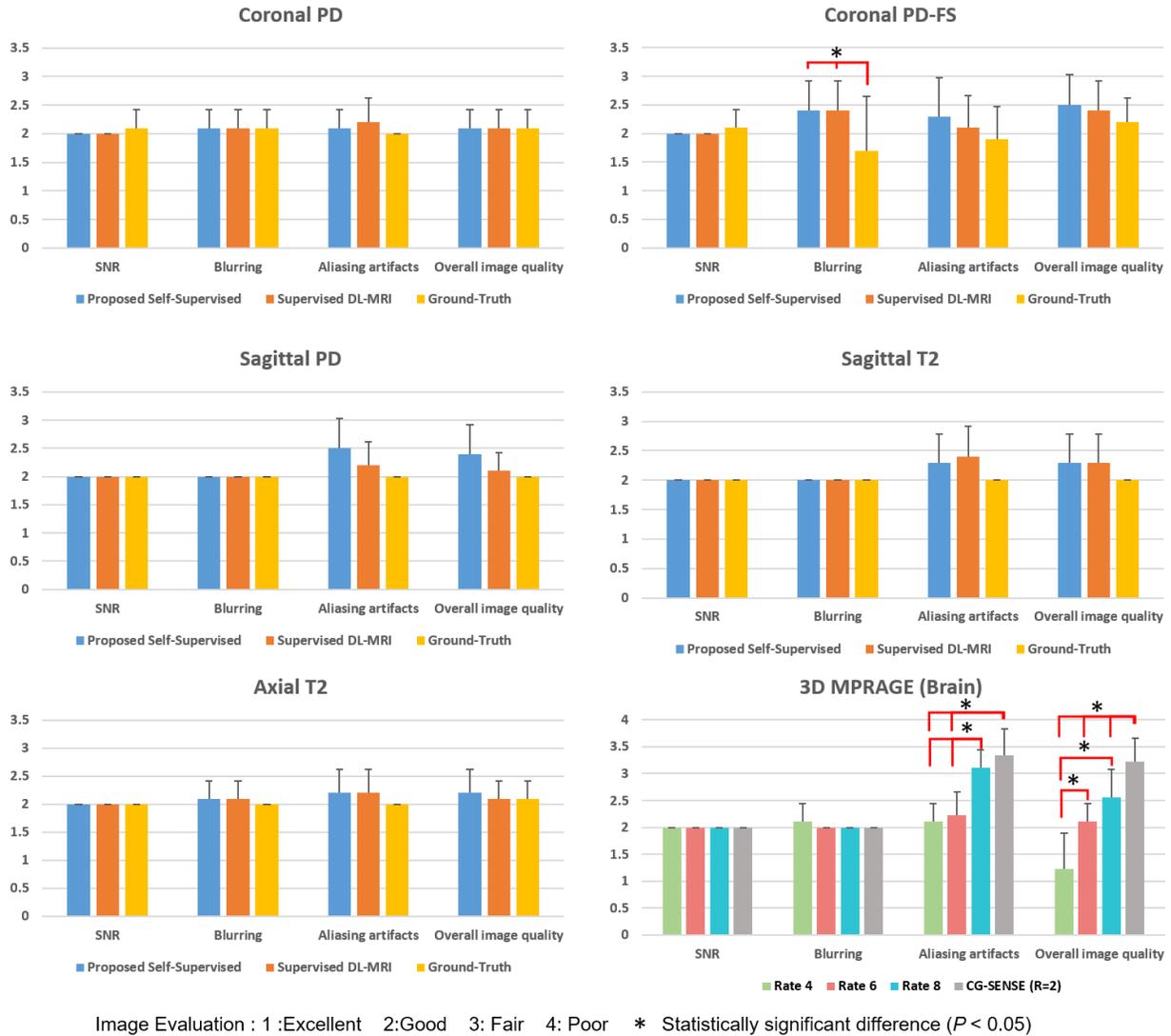

**Figure 10.** The image reading results from the clinical reader study for knee and brain datasets. Bar-plots show average reader scores and their standard deviation across the test subjects. Statistical testing was performed by one-sided Wilcoxon single-rank test, with * showing significant statistical difference with P <0.05. For knee MRI, both supervised and self-supervised DL-MRI approaches get comparable scores to the reference image in terms of SNR, blurring, aliasing artifacts and overall image quality. There was no statistical difference between reference and DL-MRI approaches in terms of the evaluation criteria for the knee datasets, except for blurring between reference and DL-MRI approaches in coronal PD-FS. For brain MRI, CG-SENSE at R =

2 and self-supervision at R = 4, 6 and 8 do not show any significant differences in terms of SNR and blurring. Self-supervision at all rates were evaluated to be significantly improved compared to CG-SENSE in terms of aliasing artifacts and overall image quality. Additionally, self-supervision at R = 6 and 8 were also significantly worse than self-supervision at R = 4 in terms of overall image quality.

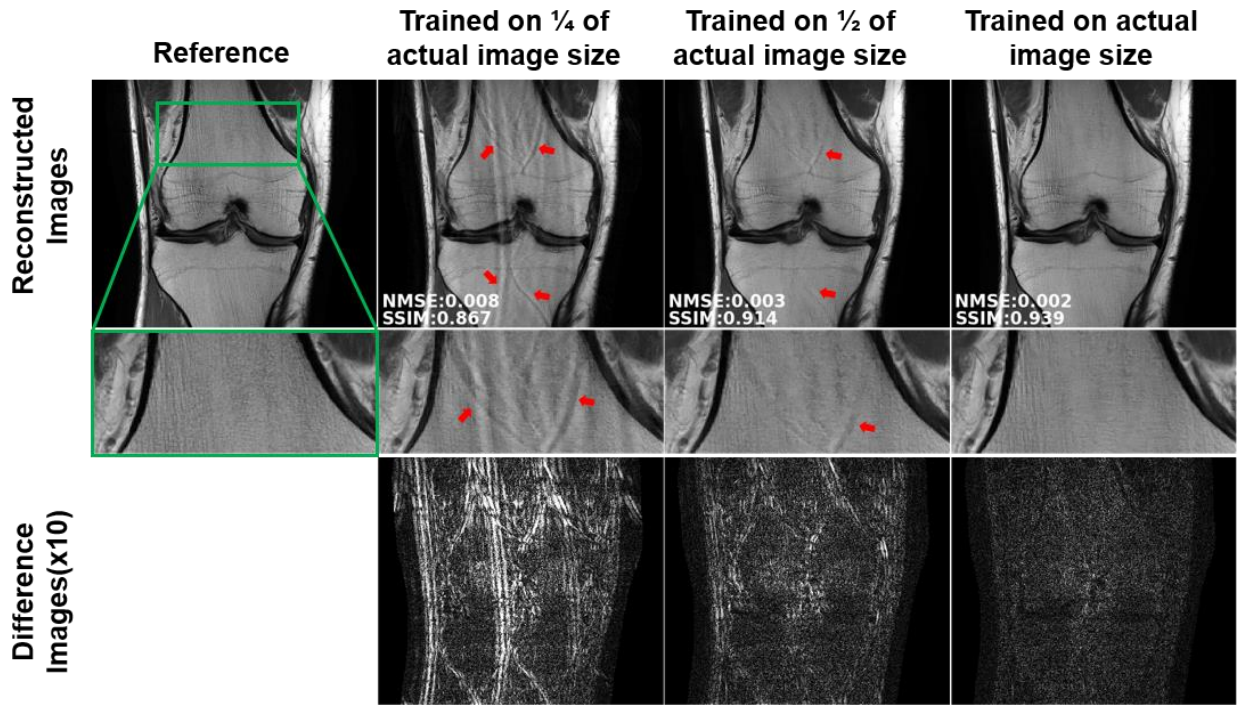

**Supporting Information Figure S1.** Reconstruction results for the generalization performance of supervised training across different image matrix sizes. The networks are trained in by taking actual k-space, the central ½ of the k-space (i.e. reducing the resolution by 2-fold), and the central ¼ of the k-space (i.e. reducing the resolution by 4-fold). All trained networks are then applied on actual size data to test generalization. The generalization performance of CNNs on actual image size degrades as training image size get smaller, with ¼ k-space performing the worst.

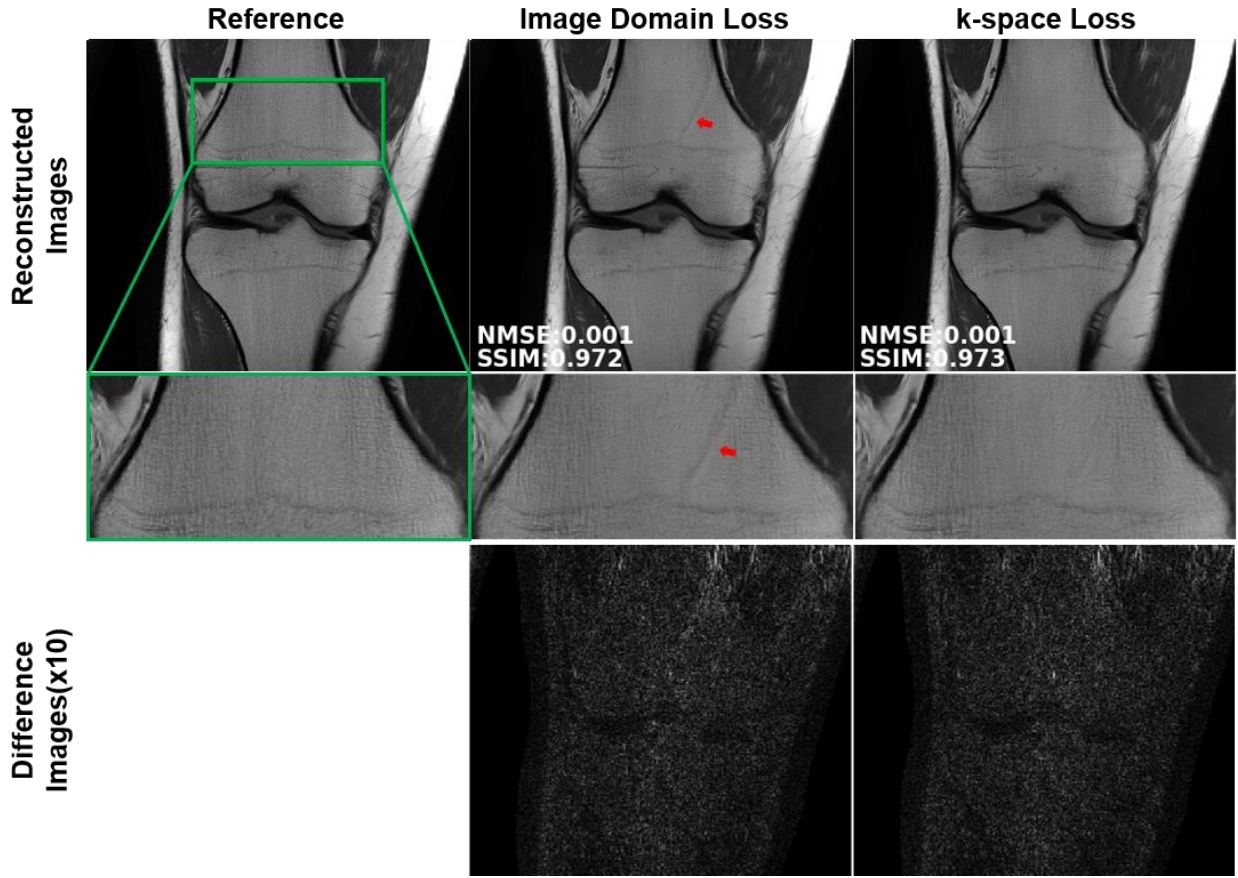

**Supporting Information Figure S2.** Reconstruction results for supervised training with image domain (Equation [7]) and k-space (Equation [8]) losses. When using image domain loss, the reconstruction suffers from residual artifacts (red arrows), whereas using k-space loss suppresses these artifacts. Difference images also show that the supervised training with k-space loss has fewer residual artifacts. Across the dataset, the two approaches perform quantitatively similar. The median and interquartile range for SSIM values across test dataset were 0.967 [0.955, 0.978], 0.966 [0.956, 0.0977], and for NMSE values were 0.001 [0.001, 0.002], 0.001 [0.001, 0.002] for supervised with image domain and k-space losses, respectively.

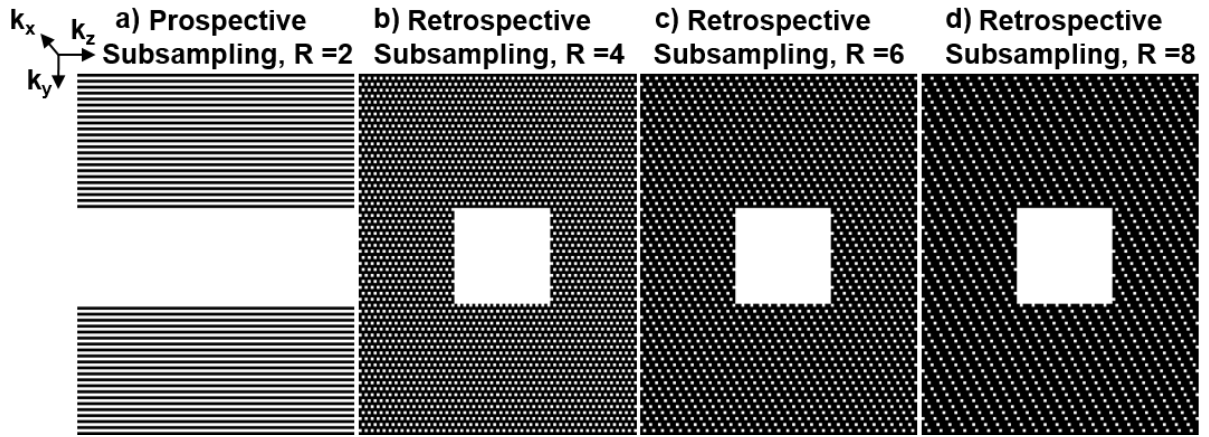

**Supporting Information Figure S3.** Sub-sampling masks used in the brain MRI study. Prospective subsampling was equispaced with R = 2 in $k_y$ and 32 ACS lines. Subsampling patterns for R = 4, 6, 8 were obtained by sheared sub-sampling, while keeping the center 32x32 ACS region in the $k_y$-$k_z$ plane.

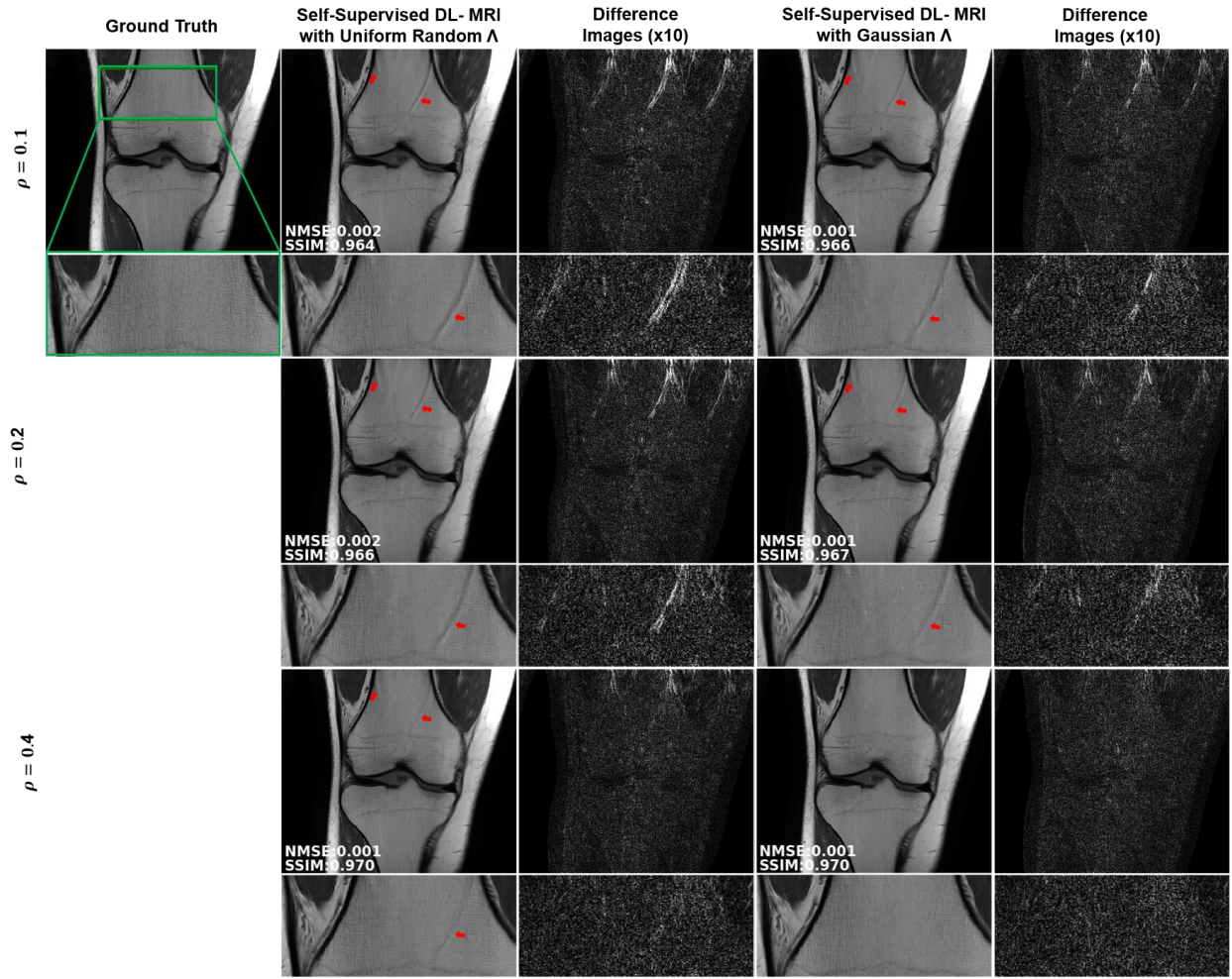

**Supporting Information Figure S4.** Reconstruction results from self-supervised training with uniform random selection and variable-density Gaussian selection of Λ for ρ ∈ 0.1, 0.2, 0.4. Gaussian random selection consistently outperforms the uniform random selection at all ρ values in terms of reconstruction quality and suppression of residual artifacts, which is also highlighted in the difference images. For ρ ∈ 0.1, 0.2 both uniform and Gaussian random selection show visible residual artifacts, marked by red arrows, with former showing more residual artifacts. For ρ = 0.4, uniform random selection still suffers from visible residual artifacts, whereas Gaussian selection further suppress those artifacts and achieves artifact free reconstruction. Difference images further confirms the observations.

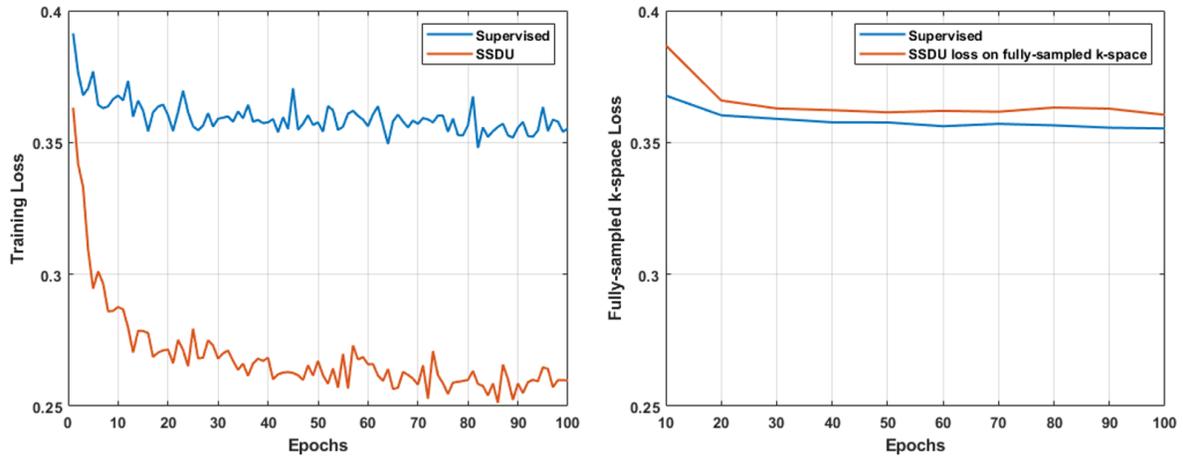

**Supporting Information Figure S5.** a) Training loss for supervised and self-supervised training approaches. In both cases, the loss decreases over epochs. Self-supervised approach achieves a lower loss value, as the loss is only measured on $\Lambda$, whereas the supervised loss is measured on the fully-sampled k-space. b) For both supervised and self-supervised training, the outputs of the networks is evaluated on the fully-sampled k-space loss, as defined in Equation [8] for every $10^{th}$ epoch. Using a similar metric, the two approaches show similar trends over epochs, with the supervised training achieving a slightly lower loss than the self-supervised approach.

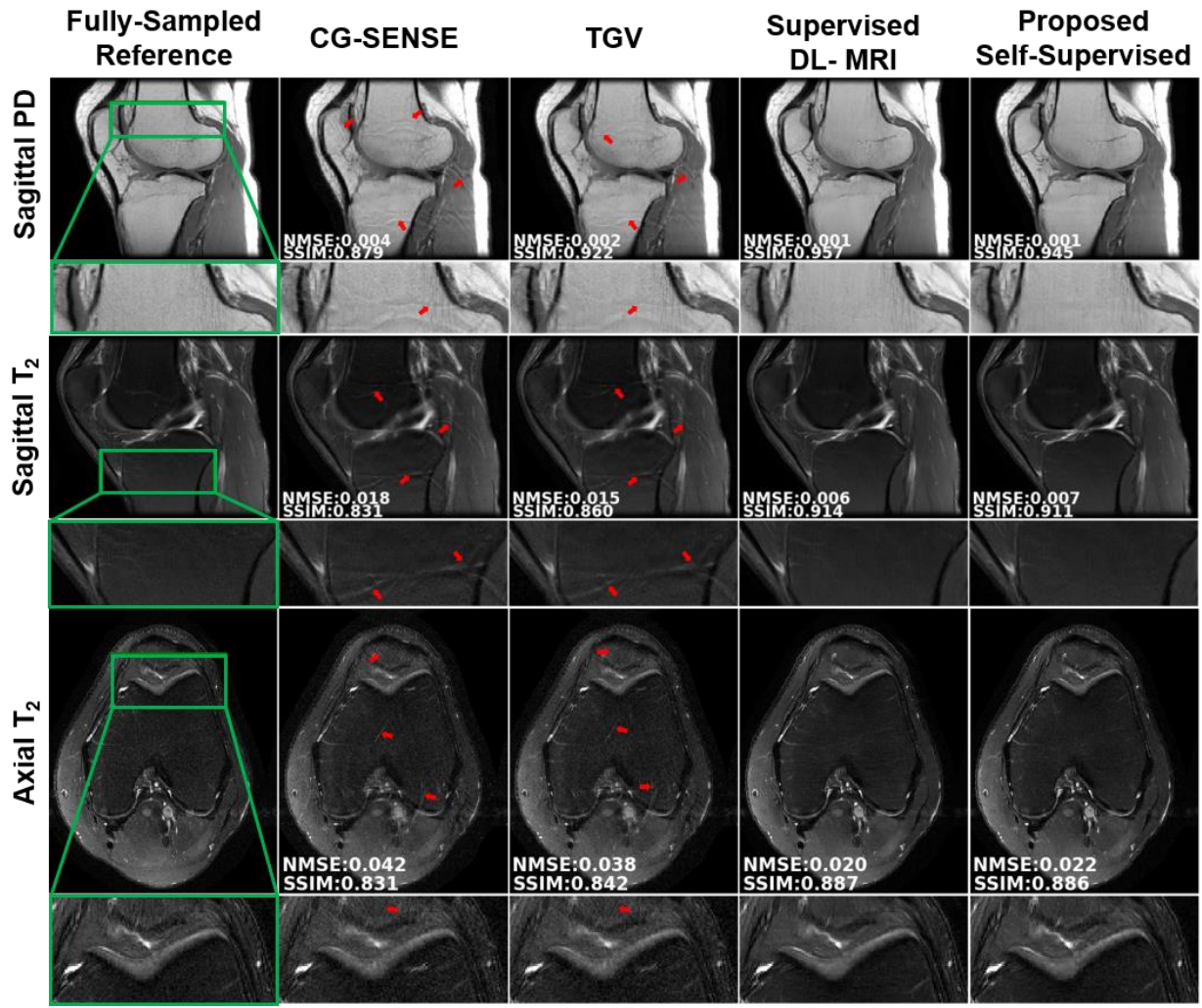

**Supporting Information Figure S6.** Representative reconstructed test slices from fastMRI sagittal PD, sagittal T₂ and axial T₂ knee sequences for retrospective equispaced undersampling R = 4. In all three sequences, CG-SENSE and TGV suffer from visible residual artifacts, marked by red arrows. Both proposed self-supervised and fully-supervised DL-MRI approaches successfully remove these residual artifacts, while showing similar quantitative and qualitative performance. Note the former does not require any fully-sampled data for training unlike the latter supervised approach.

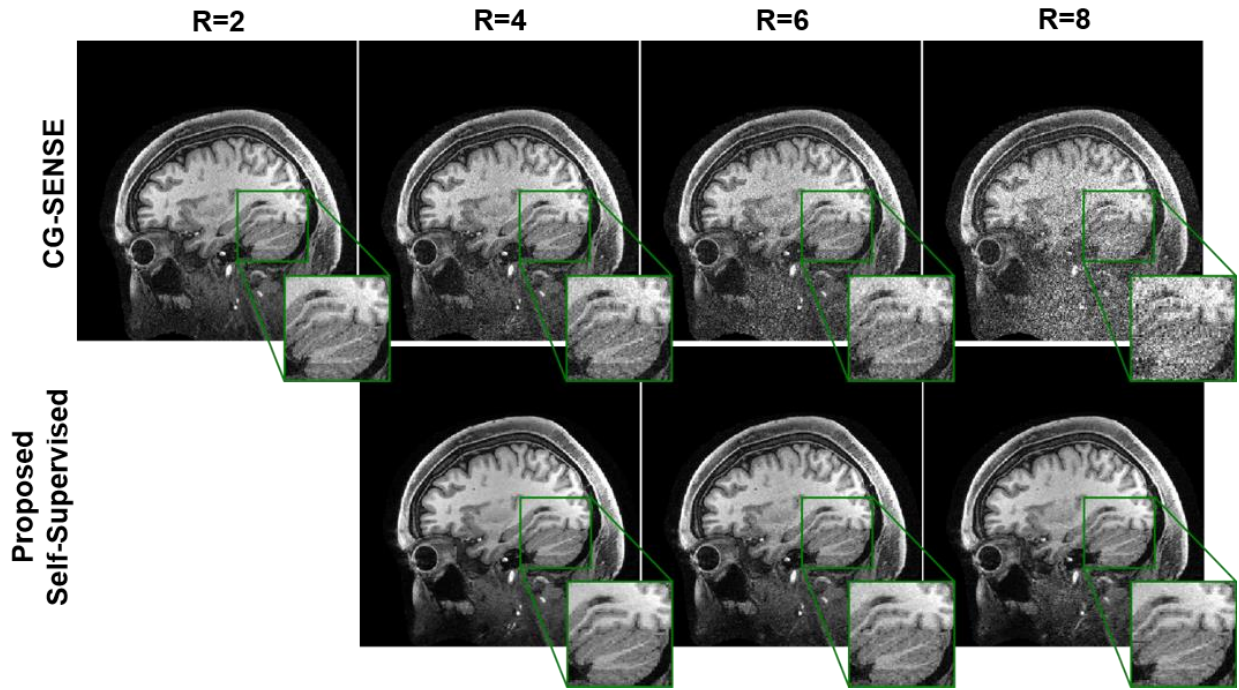

**Supporting Information Figure S7.** Reconstruction results for CG-SENSE and proposed self-supervised approach for brain MRI. CG-SENSE suffers from significant noise amplification at high acceleration rates. Proposed self-supervised approach achieves high-quality reconstruction at high acceleration rates, and achieves a lower noise amplification at rate 8 compared to CG-SENSE at acquisition acceleration rate 2.

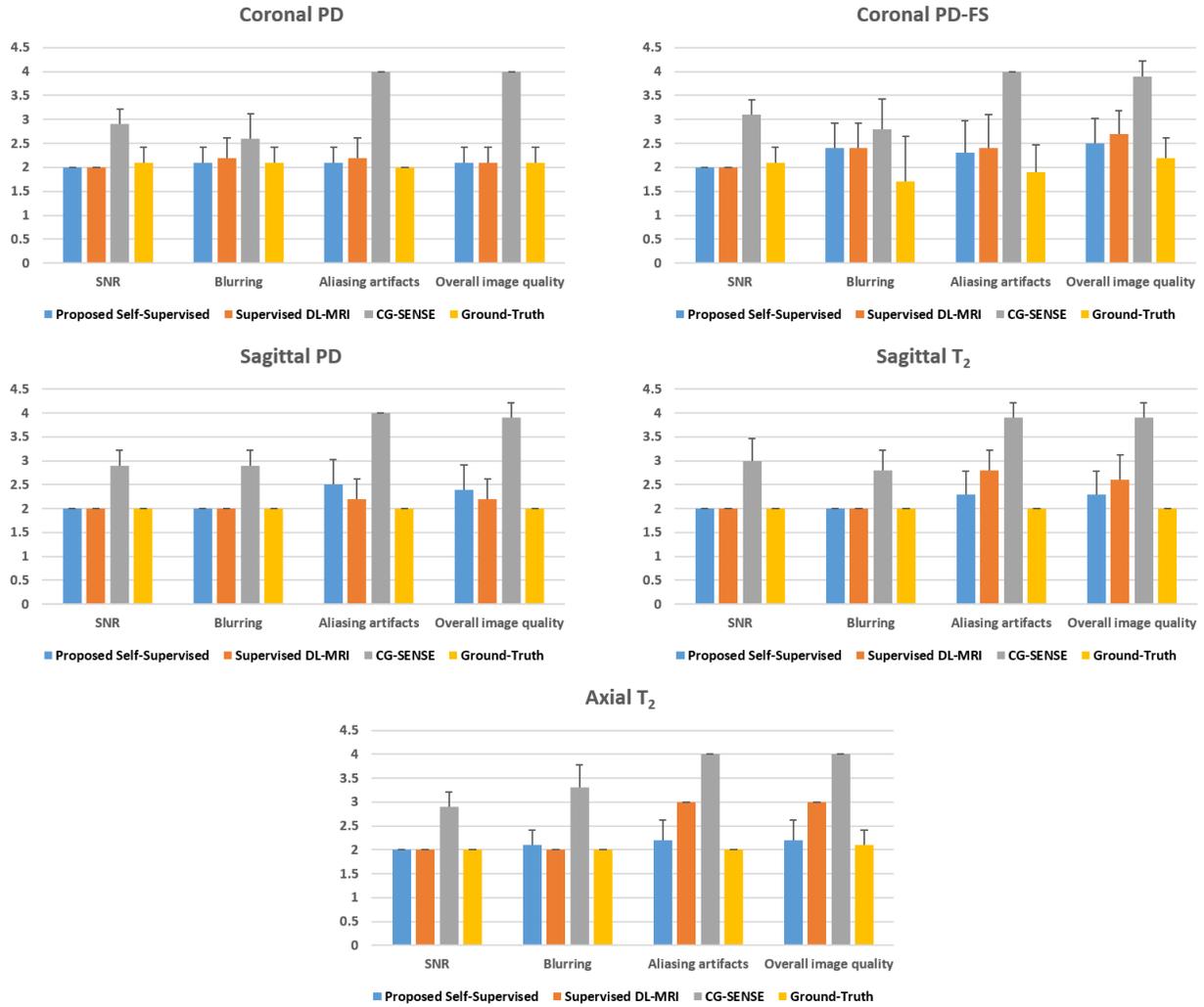

Image Evaluation : 1 :Excellent   2:Good   3: Fair   4: Poor

**Supporting Information Figure S8.** Average reader scores for all knee sequences for proposed self-supervised training, supervised training with image domain loss and CG-SENSE. Both supervised and self-supervised DL-MRI approaches get comparable scores to the reference image in terms of SNR, blurring, aliasing artifacts and overall image quality. There was no statistical difference between reference and DL-MRI approaches in terms of SNR and blurring in the knee sequences in general, except for blurring between reference and DL-MRI approaches in coronal PD-FS. In terms of aliasing artifacts and overall image quality, there were no statistical difference between reference and the two DL-MRI approaches for coronal PD, coronal PD-FS and sagittal

PD sequences. However, for sagittal $T_2$ sequence, supervised DL-MRI was ranked statistically worse than the reference, while for axial $T_2$, it was ranked lower than both the reference and self-supervised DL-MRI. Thus, in general, both DL-MRI approaches performed well, but the self-supervised approach was slightly more favored by the reader, who was blinded to the reconstruction method. CG-SENSE was significantly outperformed by both DL-MRI approaches, while showing statistically significant differences to the reference and both DL-MRI approaches for all knee sequences, except in blurring for coronal PD and PD-FS sequences. Finally, we also note that the supervised training with k-space loss (Figure 10) outperforms supervised training with image domain loss in terms of reader scores for axial $T_2$, coronal PD-FS and sagittal $T_2$ sequences.

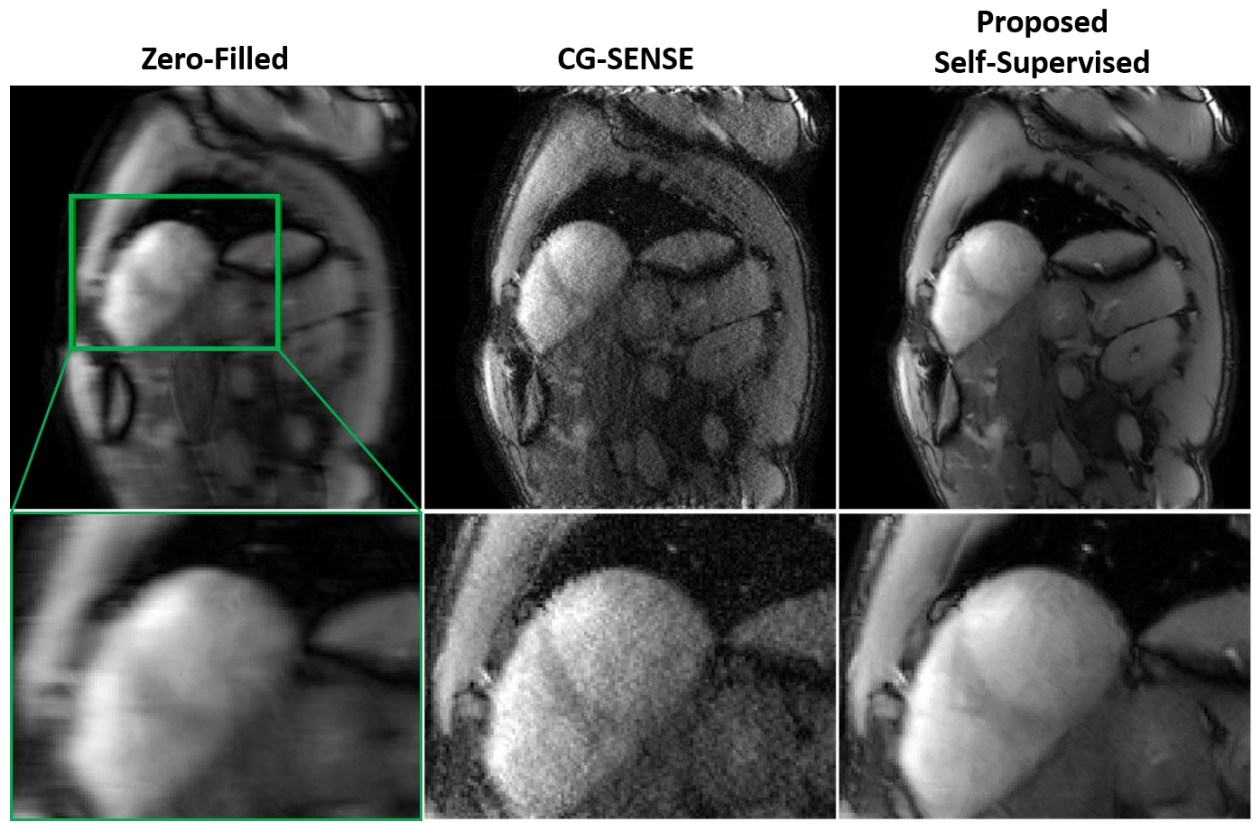

**Supporting Information Figure S9.** Reconstructed images from an 8-fold accelerated snapshot cardiac MRI data with 1.3×1.3 mm$^2$ in-plane resolution, acquired using a transient bSSFP sequence. These type of acquisitions are commonly used in cardiac parametric mapping, where the image data for one contrast weighting need to be acquired within the diastolic quiescence of one heartbeat. A fully-sampled acquisition at this higher resolution would take >700 ms, which is impossible to fit in the diastolic quiescence of a single heart-beat. Training data was acquired on 14 subjects, and testing was performed on a different subject, using the approach described in the manuscript. The proposed self-supervised approach achieves high-quality reconstruction, outperforming CG-SENSE, which suffers from residual artifacts and high noise.

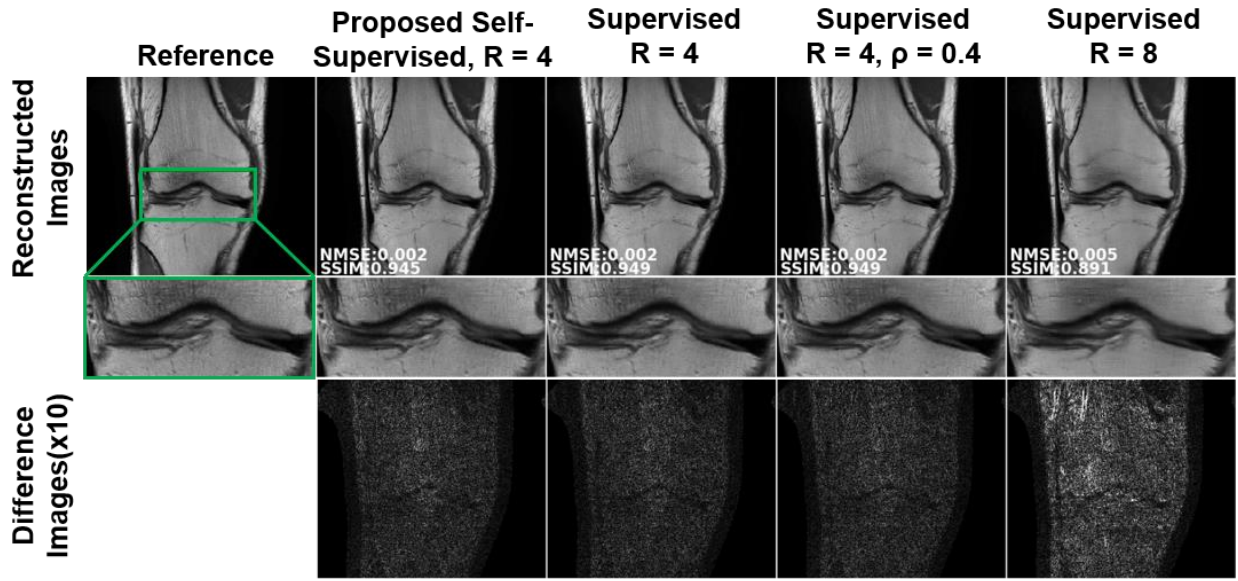

**Supporting Information Figure S10.** Reconstruction results for proposed self-supervised training at R = 4, supervised training at R = 4, R = 4 with ρ = 0.4, and R = 8. The amount of data used for self-supervised/supervised training at R = 4 (24 ACS lines) with ρ = 0.4 is 21120 k-space points, which is approximately equivalent to training the network with an equispaced undersampling pattern of R = 8 (24 ACS lines) with 21440 k-space points. The results show that supervised training at R = 4 with ρ = 0.4 is visibly similar with supervised and proposed self-supervised training at R = 4, and outperforms supervised training at R = 8. These results are visibly highlighted in difference images, which show supervised training at R = 8 suffering from residual artifacts, while other approaches show similar performance. Quantitative metrics on test dataset aligns with these qualitative assessments. The median and interquartile range for SSIM across test dataset were 0.961 [0.947, 0.972], 0.966 [0.956, 0.977], 0.966 [0.954, 0.976], 0.929 [0.908, 0.950], and NMSE were 0.002 [0.001, 0.002], 0.001 [0.001, 0.002], 0.002 [0.001, 0.002], 0.004 [0.003, 0.005] for proposed self-supervised at R = 4, supervised at R = 4, supervised at R = 4 with ρ = 0.4, and supervised at R = 8, respectively.

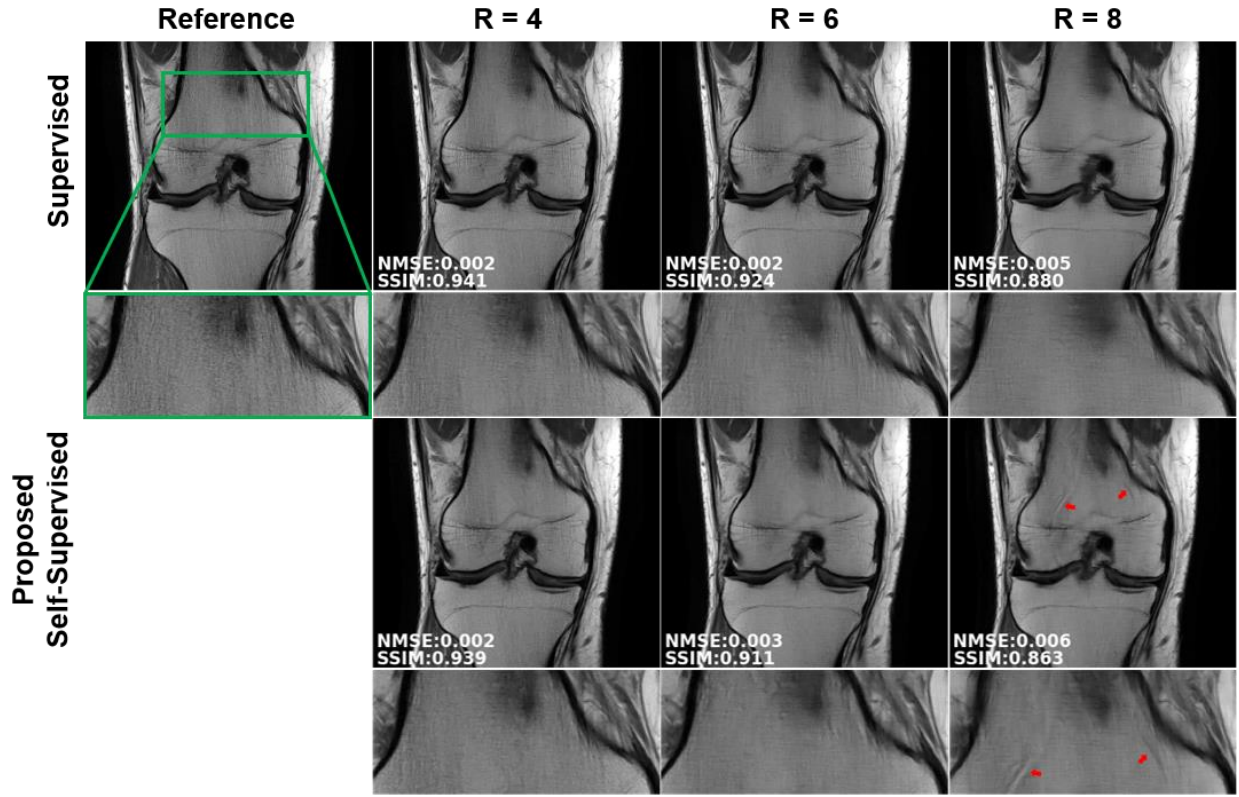

**Supporting Information Figure S11.** Reconstruction results for the coronal PD-weighted dataset at acceleration rates of 4, 6 and 8. For R = 4 and 6, the proposed self-supervised approach performs similarly with the supervised approach. However, at R = 8, the image quality degrades for both methods with more pronounced blurring, while the self-supervised approach further suffers from visible residual aliasing artifacts.

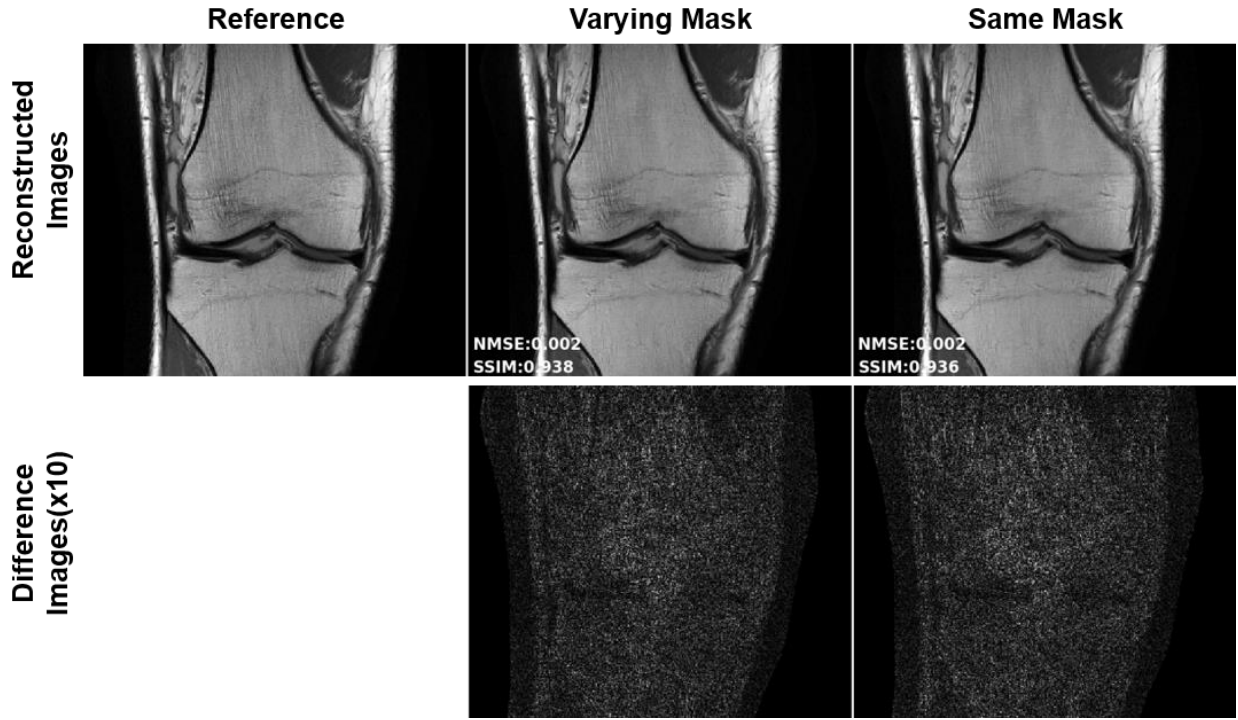

**Supporting Information Figure S12.** Reconstruction results for the proposed self-supervised approach when using same or varying sets, $\Theta$ and $\Lambda$, across different training slices. The two approaches perform similarly with the varying mask approach showing slight improvement. The median and interquartile ranges for SSIM across the test dataset were 0.959 [0.945, 0.970], 0.960 [0.947, 0.0971], and for NMSEs were 0.002 [0.001, 0.002], 0.002 [0.001, 0.002] for varying mask and same mask scenarios, respectively.

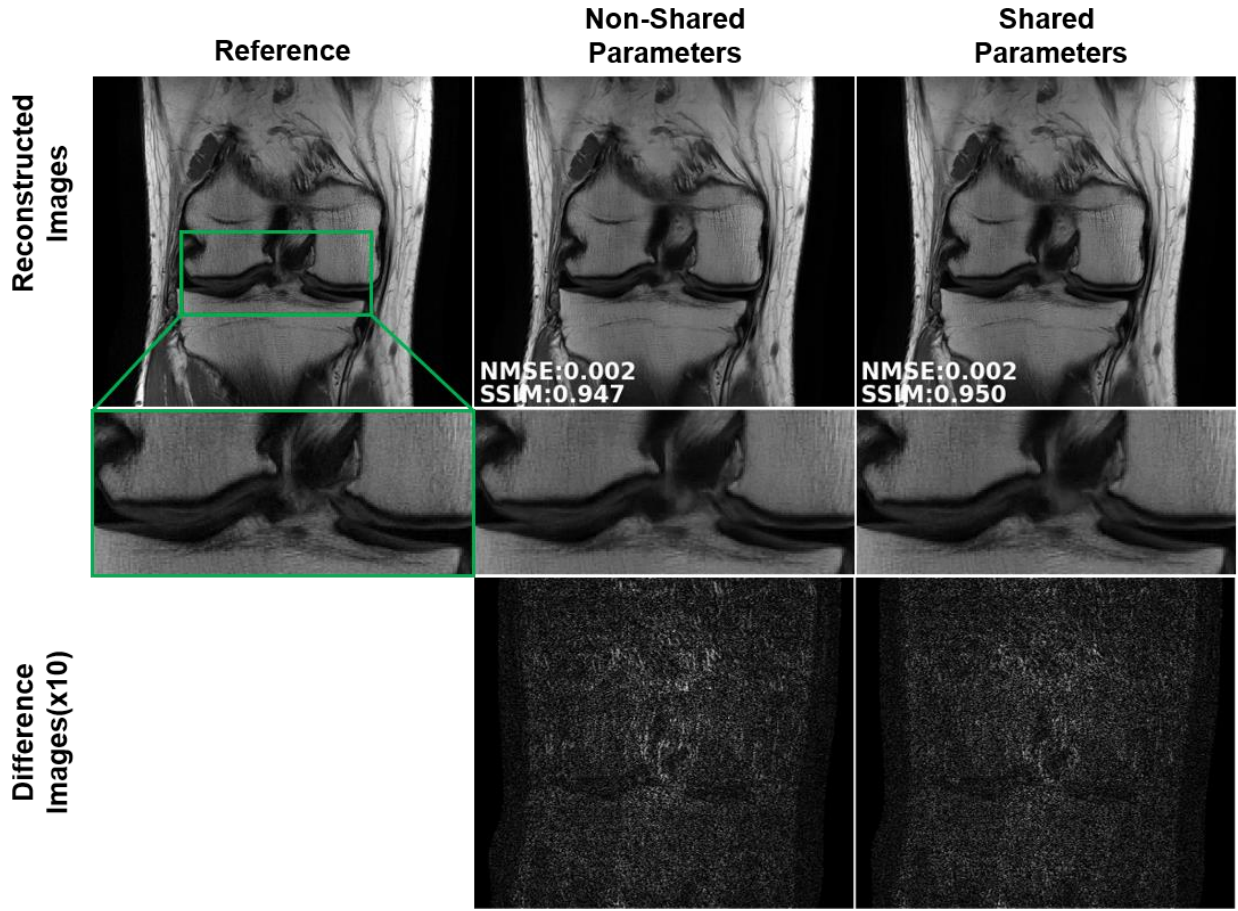

**Supporting Information Figure S13.** Reconstruction results for supervised training when using shared and distinct (non-shared) parameters across the unrolled network. The two approaches perform similarly both visually and quantitatively. The interquartile range of SSIM values across the test dataset were 0.967 [0.955, 0.978], 0.964 [0.953, 0.975], and NMSE values were 0.001 [0.001, 0.002], 0.001 [0.001, 0.002] for shared and non-shared scenarios, respectively. Note that the same training database was used for the two approaches. The non-shared approach has 10 times as many trainable parameters, and its generalization performance may benefit from a larger training database. This was not studied as it is not the focus of our study.

| Knee Sequence | TR | TE | TF | Matrix Size | In-plane Resolution | Slice Thickness | Scan time |
|---|---|---|---|---|---|---|---|
| Coronal-PD | 2750 ms | 27 ms | 4 | 320×368 | 0.49×0.44 mm$^2$ | 3 mm | 17 min |
| Coronal-PDFS | 2870 ms | 33 ms | 4 | 320×368 | 0.49×0.44 mm$^2$ | 3 mm | 18 min |
| Sagittal PD | 2800 ms | 27 ms | 4 | 384×304 | 0.46×0.36 mm$^2$ | 3 mm | 14 min |
| Sagittal T$_2$ | 4300 ms | 50 ms | 11 | 320×256 | 0.55×0.44 mm$^2$ | 3 mm | 18 min |
| Axial T$_2$ | 4000 ms | 65 ms | 9 | 320×256 | 0.55×0.44 mm$^2$ | 3 mm | 17 min |

**Supporting Information Table S1.** Imaging parameters for the knee datasets.

| Method / Metric | Disjoint sets ($\Theta= \Omega\backslash\Lambda$) | 50 % Overlap of $\Theta$ and $\Lambda$ | 100 % Overlap of $\Theta$ and $\Lambda$ ($\Omega=\Theta$) | Identical sets ($\Omega=\Theta=\Lambda$) |
|---|---|---|---|---|
| SSIM | 0.961 [0.947, 0.972] | 0.958 [0.947, 0.970] | 0.796 [0.753, 0.862] | 0.802 [0.762, 0.867] |
| NMSE | 0.002 [0.001, 0.002] | 0.002 [0.001, 0.002] | 0.009 [0.006, 0.012] | 0.009 [0.006, 0.011] |

**Supporting Information Table S2.** Median and interquartile range (25$^{th}$ -75$^{th}$ percentile) of the quantitative evaluation of SSIM and NMSE values for different overlap scenarios between $\Lambda$ and $\Theta$ when $\rho = 0.4$. Overlap %, defined as $|\Lambda\cap\Theta|/|\Lambda|$ refers to the amount of data in the loss mask $\Lambda$ that was also included in the training mask $\Theta$. Performance of the self-supervised training degrades as the amount of overlap increases.